\documentclass[aps,article,superscriptaddress]{revtex4}
\usepackage{graphics} 
\usepackage{amsmath}
\usepackage{color}
\usepackage{graphicx,epstopdf}
\DeclareMathOperator\sech{sech}

\begin{document}
\title{Pattern formation in Vlasov-Poisson plasmas beyond Landau \\
caused by the continuous spectra of electron and ion hole equilibria }

\author{Hans Schamel*}
\affiliation{Physikalisches Institut, Universit\"{a}t Bayreuth, D-95440 Bayreuth, Germany}

\date{\today}

\begin{abstract}
This review presents an upgraded wave theory adapted to the high fluctuation level of driven realistic i.e. non-idealized plasmas. Above all, this means giving up the well-known concept of a linear wave theory in favor of a thoroughly nonlinear theory. In particular, the failure to describe the formation of persistent long-lived structures by a perturbative treatment of the Vlasov equation is highlighted.
This is achieved by an extended revision  of the theory of $ \bf {stationary} $ $ \bf {coherent} $ $ \bf {waves}. $

 Based on the author's early publication (H. Schamel, Plasma Phys. 14 (1972) 905) and supported by recent Vlasov-Poisson (VP) simulations of realistic noisy plasmas, an extended framework is presented which not only covers the essential features of coherent  hole structures, but also enables one to make the necessary corrections to the current wave theory. These corrections are long overdue, in principle since 1972, and can be briefly summarized under the heading: $\bf{loss}$ $\bf{of}$ $\bf{linear}$ $\bf{Vlasov}$ $\bf{dynamics}$ when adequately addressing equilibrium states (i.e. failure of linear Landau theory and of continuous van Kampen spectra, respectively). 

In addition to the structures already known, a  number of further structures of different character are presented, including solitary electron (ion) holes with negative (positive) polarity. To each structure an evolution equation can be assigned, which governs its temporal changes. 

In contrast to the discrete phase velocities known from linear wave theories, a typical phase velocity is continuous, i.e. its pattern belongs to a multi-parametric $\bf{continuous}$ $\bf{spectrum}$ of solutions satisfying a $\bf{nonlinear}$ $\bf{ dispersion}$ $\bf{ relation}$ (NDR).
Using an NDR for continuous spectra, it is then a simple exercise to prove the existence of extremely slow solitary electron holes (SEHs).

A  linear stability analysis for single harmonic waves that successfully incorporates trapped particle effects (in contrast to previous analyses) shows an unconditional marginal stability independent of the drift between electrons and ions, which irrevocably contradicts Landau's theory.

Moreover, $\bf{holes}$ $\bf{ of}$ $\bf{ negative}$ $\bf{ energy}$ are of particular interest because they act as attractors in this dynamic system.  Due to trapping they appear an order earlier in a small-amplitude expansion scheme than in previous analyses.
Negative energy states are attained through a spontaneous acceleration of a hole that is triggered by a gap in the solution of the NDR.  Its increase in velocity is thereby accompanied by the emission of other modes such as ion sound waves raising the level of intermittent turbulence. 

The large $ \bf {diversity} $ caused by trapping means the loss of a clear identification or microscopic assignment of the parameters involved. This applies to both experimental and numerical experiments.

In summary, electrostatic structures in collisionless plasmas are determined as coherent objects by particle trapping and are therefore nonlinear, no matter how weak they are. Linear Vlasov descriptions and their perturbative nonlinear extensions, such as in the nonlinear Landau damping scenario, are unsuitable for reasons of consistency, and so fail. In order to achieve a satisfactory, if not yet complete understanding of their creation processes, a twofold paradigm shift is hence imperative:
 one from the conventional linear, discrete wave models to the nonlinear wave models dealing with $ \bf{continuous} $ $ \bf{spectra} $ $\bf{due}$ $ \bf{to} $ $ \bf{trapping} $ and a second from the BGK to the present method for the right i.e. complete handling of equilibria.
\end{abstract}


\maketitle
*email:hans.schamel@googlemail.com - \\web: www.hans-schamel.de\\

Data Availability Statement:\\
The data that supports the findings of this study are available from the corresponding author upon reasonable request.\\

\begin{flushleft}
\begin{tabular}{ |lr| } 
CONTENT\\
 \hline
I. INTRODUCTION \qquad p.2\\
\hline
II. THEORY OF ELECTRON HOLE EQUILIBRIA \qquad  p.4\\
 \hline
III. THE GALLERY OF ELEMENTARY MODES\qquad  p.8\\
\qquad III.1 The harmonic mode (single wave)\\
\qquad III.2 The privileged $\sech^4(x)$ - solitary mode\\
\qquad III.3 The Gaussian $e^{-x^2}$ - solitary mode\\
\qquad III.4  The Second Order Gaussian $e^{-\sinh^2(x)}$ - solitary mode\\
\qquad III.5  The $\sech^2(x)$ - soliton\\
\hline
IV. HOLES CAUSED BY TWO TRAPPING SCENARIOS \qquad  p.12\\
\qquad IV.1 The cnoidal electron hole (CEH)\\
\qquad IV.2 The Schamel-Korteweg-de Vries solitary electron hole (SKdV-SEH)\\
\qquad IV.3 The modified second order Gaussian SEH\\
\qquad IV.4 The undisclosed logarithmic Schamel SEH\\
\hline
V. THE CLASS OF NEGATIVELY POLARIZED SOLITARY ELECTRON HOLES (npSEHs) \qquad p.16\\
\hline
VI. THE CLASS OF ULTRA SLOW SEHs \qquad p.17\\
\hline
VII. ION TRAPPING EFFECTS AND ION HOLES\qquad p.17 \\
\qquad VII.1 Ion trapping effects\\
\qquad VII.2 Ion holes of negative and positive polarity\\
\hline
VIII. STABILITY \qquad p.19\\
\hline
IX. NEGATIVE ENERGY STATES AND SPONTANEOUS HOLE ACCELERATION \qquad p.20\\
\hline
X. TWO RELATED TOPICS: ANOMALOUS TRANSPORT AND HOLES IN SYNCHROTRONS \qquad p.22\\
\qquad X.1 Coarse grained distributions and anomalous resistivity\\
\qquad X.2 Solitary structures on hadron beams in synchrotrons\\
\hline
XI. SUMMARY AND CONCLUSIONS \qquad p.23\\
\hline
ACKNOWLEDGEMENTS \qquad p.24\\
\hline
APPENDIX A Annotated list of false statements \qquad p.24\\
APPENDIX B Derivation of (27) \qquad p.26\\
\hline
REFERENCES \qquad p.27\\
\hline


\end{tabular}
\end{flushleft}

\section{Introduction}

Firstly, the reader should be aware that they are unlikely to recognize much of what they have learned about electrostatic plasma waves so far, especially from textbooks. One reason for this  is that textbooks mainly refer to linear waves but are less communicative when it comes to  the real world of pattern formation that is strictly nonlinear without limitation. This reference to  linearity is definitely suitable for waves in the fluid description, in which higher amplitude nonlinear waves  emanate from the linear ones. In the kinetic Vlasov description, however, the connection between linear and nonlinear solution is lost due to the phase locking of the coherent structures and the associated $ \bf {trapping} $ $ \bf {nonlinearity} $, which is absent in fluid theory, but kinetically ubiquitous for structures with phase velocities that are not too high. This premise gives the description a new, largely unexplored dimension.

A second reason is that in the past the wrong method was preferred by the community in the nonlinear regime, namely the BGK method \cite{BGK}. This method has definitely historical merits as it was the first time that a correct Vlasov-Poisson (VP) solution could be obtained by introducing the trapped particle concept. However, as is explained in more detail also later, the BGK method cannot provide a complete solution, since the phase velocity, the second part of a nonlinear solution of not less importance,  remains indefinite. A correct phase velocity is, for example, necessary to set up the decisive evolution equation or to decide on the predominant wave energy. In addition, the shape of the electrical wave potential $ \phi (x) $, which is a prerequisite for handling the BGK method, can no longer be specified mathematically for a typical solution namely when more than one trapping scenario is involved.

 Linear theory is thus reserved and applicable for specially prepared, calm plasmas. The first experimental verification of Landau / Langmuir waves by Derfler and Simonen \cite{DeSi66}, by measuring of the Bohm-Gross dispersion and the damping rate, for example, could only be carried out successfully after they had painstakingly \cite{DERFLER} created the prerequisites for the validity of the Landau theory, namely a quiet background plasma and  a perturbation that satisfies the "topological constraint" $|\partial_v f_1|<<|\partial_v f_0|$ valid at every moment of evolution. For the "nonlinear Landau damping" (NLD) scenario linear theory only applies in the early phase of evolution, i.e. before saturation on a much lower, but $\bf{nonlinear}$ level \cite{Manfredi97}. 
To the surprise of many, the structure is nonlinear in this late, lowest energy state.
Note that this latter, dynamically calmer state is absent from the perturbation analysis by Mouhot and Villani \cite {MV11, V14, Cit7} since trapping effects are neglected by them. The scenario of the NLD is hence only completely solved if coherent nonlinear structures, as we will develop in this article, are included, even if Landau's prerequisites apply initially.
Therefore, to achieve consistency in the NLD scenario, it is imperative to consider trapping.\\

If the second, the topological condition is violated, the damping can be very different or even missing \cite {KS96a}.\\

In \cite {BS92}, to present a second well-known example, the two-stream instability, the early phase of linearly dominated, but rather violent nonlinear development (described by mode coupling, including mode slaving and the tendency to wave collapse) is replaced by a sudden  calming and saturation of the evolution through particle trapping.
This calm phase in the structure formation caused by trapping, to say it again, is our concern in a general context beyond the Landau scenario.\\
In general, linear wave theory describes pretty well incoherent waves of small amplitudes and random phases but has no chance of meeting the abundance of coherent structures that establish in driven, noisy plamas triggered for example by seeds or eddies. By localized seeds particle trapping is involved from the very beginning and an a priori linearization of the VP system is no longer useful. The Landau theory is therefore not suitable for describing pattern formation caused by seeds.

The correct view, therefore,  is  that the Vlasov equation, as a nonlinear equation, must first be solved before the small amplitude limit is taken, and not the other way around. In other words: The smallness of a wave has to be seen as a limiting case of the nonlinear solution and not by solving a wrong equation,  the linearly truncated Vlasov equation.The good solvability of the linear Vlasov equation does not necessarily offer a valid ticket to the realm of nonlinear structures. \\
In the current-driven plasma situation, this premise is justified by comparing both solutions, the linear and the nonlinear.
In Fig. 2 of \cite{S12}, in which the two distributions are compared with one another in the resonant region, the differences are clearly visible. While the nonlinear solution behaves well, the linear solution involves principal value and delta function singularities in the van Kampen case or manipulations of the background distribution(s) at resonance in the Landau case that should mimic trapped particles. These manipulations are artificial, i.e. not carried out correctly to the end and hence miss nonlinear self-consistency. They hence lack mathematical seriousness and rigor. These differences are retained and do not disappear in the infinitesimal amplitude limit. We will address this point again in Section III.1.\\

The main goal of the present paper therefore is to provide the reader with the necessary components of a correct nonlinear wave theory.\\

In Appendix A  we list and comment some of the well known but inappropriate and outdated statements that  do not stand up to critical analysis mostly because of their linear origin. \\

From this fake news we now come to the indisputable facts. The following applies in general: \\

 (i) Vlasov-Poisson (VP) structural equilibria are strictly nonlinear.\\
This means that any linear approximation is microscopically doomed to failure.\\

 (ii) They are represented by an infinite variety. \\
This implies that their wave potential $ \phi (x-v_0t) $ is typically undisclosed, i.e. it can no longer be expressed using mathematically known functions. \\

 (iii) They belong to a continuous spectrum.\\
This results from the fact that $ v_0 $ is a solution of a nonlinear dispersion relation (NDR) that includes in a self-consistent manner the trapping  nonlinearity (TN) and generally involves the $\gamma$ trapping scenario. \\

(iv)  Schamel's pseudo-potential method provides the most adequate description .\\
It offers a complete and consistent solution in terms of shape $\phi(x)$, phase velocity $v_0$ and distribution functions.\\

(v) On the pseudo-potential level there is a kind of $ \bf {nonlinear} $ $ \bf {superposition} $ $ \bf {principle} $.\\
This means that one can find  new solutions by linearly superimposing two or more pseudo-potentials (as demonstrated excessively in this article).\\

Ad (ii): From a microscopic point of view, there are innumerable possibilities to create a given, prescribed macroscopic structure together with its phase velocity by letting different trapping scenarios share. But note that the situation is even worse. Even with an additional measurement of $ f_ {et} $, a clear identification of a structure is not possible, since the solutions are too close to one another to be experimentally differentiable. Due to the errors and inaccuracies of the measurement and of the numerical procedure they can no longer be resolved. This applies to all measurements: in the experiments and in the numerics. \\

Ad (iv): Half a century ago, the current author presented a complete and consistent solution for VP equilibria \cite{S72}. His method is first to look for a complete solution to the Vlasov equation(s) and then, in a second step, to solve Poisson's equation thus ensuring self-consistency. A major advance was that, as physically required, the structure was embedded in a plasma and therefore the undisturbed plasma background  came correctly out in the limit of a disappearing wave. The method inevitably provided the correct nonlinear dispersion relation and the pseudo-potential in canonical form, components of a complete wave theory. Moreover,  the normalizations of the distributions were correctly taken into account for the first time. Some of these innovations were rediscovered a decade or nearly two decades later by respected theoreticians, who touted their findings as supposedly new and groundbreaking (keywords: hole theory and electron acoustic mode, respectively).\\

 To mention some more innovations, the work in \cite{S72} already contains wave solutions of finite amplitude and their correct small amplitude limit, including non-isothermal ionic acoustic solitary waves. In addition, the second dispersion branch, which was later termed  "slow ion acoustic wave" (SIAW) branch and which is e.g. part of the  thumb-teardrop dispersion relation \cite{ TG18, S19}, was advertised for the first time in this article (see the Figs. 2.3 and eqs. (32a,b) of \cite{S72}). In the small amplitude limit, also cnoidal (his terminology was snoidal) wave solutions including Schamel's evolution equation of the Korteweg-de Vries type were presented.\\
Later the first electron hole solution of the Vlasov-Poisson system  was  presented in \cite{S79}, which provided the first intrinsically correct description for the holes measured experimentally e.g. by \cite{Saeki79}. A brief historical review of the electron hole theory can be found in Appendix A of \cite{S20a}. The first basic articles on ion holes or double layers can be found in  \cite{SB80, BS81} and \cite{SB83}, respectively. Two early reviews were presented in \cite{S82,S86}. A drift between electrons and ions was first considered in \cite{SM94}. \\

In this article, the theory of electron-hole equilibria is unfolded in detail, with an emphasis on its occurrence in collision-free, current-carrying, noisy plasmas. It offers new insights into the dynamics of holes triggered by tiny seeds particularly in linearly subcritical plasmas as seen in the highly accurate numerical simulations of Mandal $\&$ Sharma  \cite{SMS17,MSS18,MSS20,SMS20a,SMS20b}. It explains why a hole is suddenly accelerated during its evolution and why it settles on the high energy tail of the distribution where the slope is negative, rather than on the low energy, positively inclined tail, i.e. between the ion and the electron peak,  as one would expect from a linear perspective. The existence of privileged electron holes, which exist as nonlinear structures up to the infinitesimal amplitude limit, is discussed in detail and further simplified  modes are recovered. 
The appearance of intrinsic substructures in the trapped particle distribution and in the macroscopic particle densities are further new elements that can be understood as well \cite{SMS17,MSS18,MSS20,SMS20a,SMS20b}. Several new solitary wave types are presented and it is proved that the majority of possible solitary wave solutions refer to mathematically undisclosed potentials $\phi(x)$ \cite{S20a,S20b}. Finally, the negative energy concept associated with these modes offers a new avenue of plasma instability triggered by tiny seeds.

\section{Theory of electron hole equilibria }
To describe the theory as transparently as possible we study in a first step to a two-component, current-driven plasma in which trapping effects refer only to the electrons, i.e. we focus firstly on electron trapping effects for electron holes (EHs) propagating in the electron thermal range. In order not to appear too inflated, ions are allowed to be mobile, but without ion trapping (reflection) effects, which are included in a second step later. Instead, we want to get to know the influence of electron trapping as well as possible.The wavelength of the structure is arbitrary at the beginning, but is later assumed to be infinite in the solitary wave limit. \\

As said, the Schamel method consists in first solving the Vlasov equation before taking the small amplitude limit, and not vice versa. We therefore start with a stationary solution of the full electron Vlasov equation, which reads in the wave frame where the structure is at rest:  $(v\partial_x + \phi'(x) \partial_v)f_e(x,v)=0$. It is solved by any function of the single particle energy $\varepsilon:=\frac{v^2}{2}-\phi(x)$ valid for the whole velocity range. For free particles there is another (discrete) constant of motion, the sign of the velocity $\sigma:=v/|v|$, which is needed for traveling holes having a nonzero phase velocity $v_0$. This, together with the requirement that the electrons wlog obey a shifted Maxwellian in the undisturbed case, results in the following Schamel distribution \cite{S72,S00,S20a,S20b} :\\
  
\begin{eqnarray}\nonumber
  f_e(x,v)=\frac{1+k_0^2 \psi/2}{\sqrt{2\pi}} \biggl(\theta(\varepsilon) e ^{- (\sigma \sqrt{2\varepsilon} - \tilde v_D)^2/2} + \\
\theta(-\varepsilon) e^{- \tilde  v_D^2/2} \{1 + [\gamma +\chi_1 \ln(-\varepsilon) +\chi_2 \ln(-\varepsilon)^2] (-\varepsilon)^{1/2} -\beta \varepsilon +\zeta (-\varepsilon)^{3/2} \} \biggr),
\label{f-schamel}
\end{eqnarray}

where the curled bracket in  (1):
  $ \{1 + [\gamma +\chi_1 \ln(-\varepsilon) +\chi_2 \ln(-\varepsilon)^2] (-\varepsilon)^{1/2} -\beta \varepsilon +\zeta (-\varepsilon)^{3/2}\} $   represents the contributions of the trapping scenarios under consideration.
In this equation $\theta(x)$ represents the Heavyside step function. 
We use normalized quantities such that the velocity is normalized by the (unperturbed) electron thermal velocity, the electron potential energy by the electron thermal energy, and the space by the Debye length.\\

It  results from the Galileian shift $\tilde v_D$ of the Maxwellian given  in the unperturbed case by   $f_M(v)=\frac{1}{\sqrt{2\pi}} e^{- (v- \tilde v_D)^2/2}$ and from the replacement of $v$ by $\sigma \sqrt{2\varepsilon}$ as an effect of the perturbation. This holds for $\varepsilon>0$, which represents the free electron region. The gap in between $\sigma>0$ and $\sigma<0$ , when $\varepsilon\le0$, refers to trapped electrons.
The distribution $f_e(x,v)$  is thus a function of the two constants of motion, 
$\epsilon$ and $\sigma$,  and consists of two parts, the contribution of untrapped particles, $\varepsilon>0$, and the one of trapped particles, $\varepsilon \le 0$. Trapping is therewith controled  by the five parameters $\gamma$, $\beta$, $\zeta$, $\chi_1$, $\chi_2$, the first three refer to a perturbative treatment of trapped particle effects and represent the first three elements of a Taylor expansion with respect to $\sqrt{-\varepsilon}$ of a more general, exponential  $f_{et}$, whereas the fourth and fifth, $\chi_1$ and $\chi_2$ ,  are definitely non-perturbative in nature.
Note that $f_e(x,v)$ is continuous across the separatrix and it is assumed that $0 \le \phi(x) \le\psi<<1$. \\
The electron density $n_e(\phi)$ is obtained by a velocity integration.\\
We mention in passing that studies of finite amplitude,  $\psi \simeq O(1)$, electron holes and strong double layers \cite{BS81,SB83} use a similar but unexpanded distribution of trapped electrons in which case Schamel's functions  $\mathcal{K}(x,y), \mathcal{H}(x,a,b)$, defined e.g. in \cite{S72,BS81,S82, DasS05}, are involved.
  Here we restrict our analysis to weak solutions, $\psi<<1$. It can either be done by the velocity integration of (1) first and a subsequent Taylor expansion, using $\phi<<1$, as done e.g. in \cite{S72,S79,S82,S86} or by the Taylor expansion of (1) first, followed by the velocity integration, as done e.g. in \cite{S73,S75,KS96a,S20b}. Both cases yield the same result:

\begin{eqnarray}
n_e(\phi)=(1 +\frac{1}{2}k_0^2\psi)\biggl[1+\biggl( A
  -\frac{1}{2} Z_r'(\frac{ \tilde v_D}{\sqrt 2})  -\frac{5B}{4\sqrt \psi}\sqrt \phi  + C \phi + (D_1 +a_1D_2)  \ln \phi + D_2  \ln^2\phi   \biggr) \phi  + \frac{1}{16} Z_r'''(\frac{ \tilde v_D}{\sqrt 2}) \phi^2 + ...\biggr]
\end{eqnarray}
where $A:=(\Gamma +\frac{a_1}{2}D_1 +a_2D_2)$,
$B:=\frac{16}{15} b(\beta,  \tilde v_D) \sqrt \psi$  with $b(\beta, \tilde v_D):= \frac{1}{\sqrt \pi} (1- \beta -  \tilde v_D^2)e^{- \tilde v_D^2/2}$ and \\$(\Gamma, C, D_1, D_2):=\frac{ \sqrt \pi}{2} e^{-\frac{\tilde v_D^2}{2}}(\gamma, \frac{3\zeta}{4}, \chi_1, \chi_2)$.
The constants are given by $a_1= 2(1-2\ln2)=-0.773$ and $a_2=-2+\ln4(\ln4-2)+\pi^2/3    =0.439$. The other quantities are defined by \\

$\tilde v_D:=v_D - v_0, \qquad u_0:=\sqrt{\frac{T_e m_i}{T_i m_e}} v_0,  \qquad \theta:=\frac{T_e}{T_i}$.\\

Note that all trapping parameters $(A,B,C,D_1,D_2 ; \Gamma)$ carry the factor $e^{- \tilde v_D^2/2}$, i.e. they vanish in the large $|\tilde v_D|$ limit.They therefore only influence the pattern formation for moderate and small values of $|\tilde v_D|$ . As expected, their influence on high-speed Langmuir waves is therefore negligible.\\

Before we go any further, let's secure this density expression by referring to known special cases.\\

In case of $k_0=0$,  $\tilde v_D$=0 and of zero trapping parameters we find by utilizing $-\frac{1}{2}Z'_r(0)=1$ and  $-\frac{1}{2}Z'''_r(0)=-4$ the well-known Boltzmann expression for $n_e(\phi)$: $n_e(\phi)= 1 + \phi + \frac{1}{2} \phi^2 + ..$. This particularly confirms the last term in (2) contrary to a different statement found in the literature \cite{Hu17}.\\
If we keep ($k_0$,  $\tilde v_D$, $B$) but neglect ($\Gamma$,$C$, $D_1$,$D_2$) we get \\
$n_e(\phi)=(1 +\frac{1}{2}k_0^2\psi)\biggl[1 -\frac{1}{2} Z_r'(\frac{ \tilde v_D}{\sqrt 2})\phi  -\frac{5B}{4\sqrt \psi} \phi^{3/2}  + \frac{1}{16} Z_r'''(\frac{ \tilde v_D}{\sqrt 2}) \phi^2 +...\biggr]$\\
which is identical with (3.9) of \cite{KS96a}.    
And last but not least, if we just keep  $(\Gamma,D_1,D_2)$ as non-zero, we get\\
$n_e(\phi)=\biggl[1+\biggl( A
  -\frac{1}{2} Z_r'(\frac{ \tilde v_D}{\sqrt 2})  + (D_1 +a_1D_2)  \ln \phi + D_2  \ln^2\phi   \biggr) \phi  + ...\biggr]$,\\
which is the expression (2) of \cite{S20a}.\\

For the ion density we take an expression that incorporates  $O(\psi^2)$ terms but  neglect ion trapping effects and refer to a straightforward  extension that reduces to the known expressions in limiting cases:
\begin{eqnarray}
n_i(\phi)= 1 + \frac{\theta \phi}{2} Z_r'(\frac{ u_0}{\sqrt 2}) + \frac{\theta^2 \phi^2}{16}Z_r'''(\frac{u_0}{\sqrt 2})    +...
\end{eqnarray}
It reduces in the $u_0 \rightarrow 0$ limit to
$n_i=1 -\theta\phi +\frac{(\theta \phi)^2}{2} \approx e^{-\theta \phi}$, the expected Boltzmann value. On the other hand, in the "cold ion" or large $u_0$ limit we receive $n_i=1 +\frac{\theta\phi}{u_0^2} +\frac{3(\theta \phi)^2}{2u_0^4} \approx  \frac{1}{\sqrt{1-\frac{2\theta \phi}{u_0^2}}}$ valid under the constraint $|\theta \phi/u_0^2|<<1$.\\
Later in Section VII we will turn to a more general ion density that also includes ion trapping effects.
Notice that the $\bf{immobile}$ ion case is automatically included in  $n_i$, namely by setting $\theta=0$.\\

It should be emphasized that these density expressions are permissible since they are derived from solutions of the Vlasov equation. This is in contrast to publications where the $\phi$ - dependence is simply imposed without guaranteeing that a valid distribution, especially that for free and trapped particles,  stands behind. As long as this justification is lacking, these publications remain unfounded and tend to build castles in the air instead of delivering a proper theory \cite{Cairns95, MC96, Guio03}. \\
As said, in case of finite amplitudes and Maxwellian plasmas Schamel's functions  $\mathcal{K}(x,y), \mathcal{H}(x,a,b)$ \cite{S72,BS81,S82,SB83, DasS05, Goswami08} are involved, for nonextensive distributions, however, such as $\kappa$-distributions (e.g. \cite{Tribeche12}), a corresponding extension is still missing.\\

After insertion of the densities (2) and (3) into Poisson's equation, $\phi''(x) = n_e(\phi) -  n_i(\phi)=:- \mathcal V'(\phi)$, where in the last step the pseudo-potential $\mathcal V(\phi)$ has been introduced, we get (ignoring a term of $O(\psi^2)$ connected with $k_0^2$)
\begin{eqnarray}\nonumber
-\mathcal{V}'(\phi)=\frac{k_0^2\psi}{2} +\phi \biggl[ \biggl(A
  -\frac{1}{2} Z_r'(\frac{ \tilde v_D}{\sqrt 2}) - \frac{\theta}{2} Z_r'(\frac{ u_0}{\sqrt 2}) \biggr) -\frac{5B}{4\sqrt \psi} \phi^{1/2} + C\phi + (D_1 + a_1D_2) \ln \phi + D_2 \ln^2\phi +\\
 \frac{1}{16}[ Z_r'''(\frac{ \tilde v_D}{\sqrt 2}) -\theta^2 Z_r'''(\frac{ u_0}{\sqrt 2})]\phi\biggr]
\end{eqnarray}
and by integration with $\mathcal{V}(0)=0$

\begin{eqnarray}
-\mathcal{V}(\phi)=\frac{k_0^2 \phi \psi}{2} +\frac{\phi^2}{2}\biggl[\biggl( A
  -\frac{1}{2} Z_r'(\frac{ \tilde v_D}{\sqrt 2}) - \frac{\theta}{2} Z_r'(\frac{ u_0}{\sqrt 2})  \biggr) - B\sqrt\frac{\phi}{\psi}  + [D_1 +(a_1-1)D_2] (-\frac{1}{2} + \ln \phi) + D_2 \ln^2\phi + \tilde C \phi \biggr]
\end{eqnarray}
which we abbreviate as: $-\mathcal{V}_0(\phi)$ since it is used only temporarily.
In (5) we have also introduced the quantity $\tilde C$ which is defined by
 $\tilde C:= \frac{2C}{3} +\frac{1}{24}[ Z_r'''(\frac{ \tilde v_D}{\sqrt 2}) -\theta^2 Z_r'''(\frac{ u_0}{\sqrt 2})]$.\\
(5) reduces to the known expression (4) of \cite{S20b} in case of immobile ions ( $\theta=0$), of ($D_1=D,D_2=0$) and of a negligible $O(\phi^3)$ term.\\

 The necessary constraint of a second zero of $\mathcal V_0(\phi)$, at $\phi=\psi$, yields 
\begin{eqnarray}
k_0^2 +\biggl( A -\frac{1}{2} Z_r'(\frac{ \tilde v_D}{\sqrt 2}) - \frac{\theta}{2} Z_r'(\frac{ u_0}{\sqrt 2}) \biggr ) - B +\biggr[D_1 + (a_1-1)D_2](-\frac{1}{2} +\ln \psi) + D_2\ln^2\psi + \tilde C \psi \biggr]=0.
\end{eqnarray}

This expression is identical with (5) of \cite{S20a} and (5) of \cite{S20b} in the appropriate limits. It represents the equation for determining the phase velocity $v_0$ as a function of the other parameters and is hence the nonlinear dispersion relation (NDR), a relation of emminent importance. \\

Replacing the first big bracket of (5) by (6) we get:

\begin{eqnarray}
-\mathcal {V}(\phi)= \frac{k_0^2}{2} \phi(\psi-\phi) +     \frac{\phi^2}{2}\biggl[   B (1- \sqrt{\frac{\phi}{\psi}})  +    (D_1-rD_2)\ln\frac{\phi}{\psi}  + D_2 \ln^2(\frac{\phi}{\psi}) - \tilde C \psi (1-\frac{\phi}{\psi}) \biggr]
\end{eqnarray}

where $r$ is $r:=1-a_1 -2\ln\psi=1.773 - 2\ln\psi$. In this form $\mathcal {V}(\phi)$ automatically satisfies $\mathcal {V}(\psi)=0$, a form we will call $\bf{canonical}$. \\

To obtain finally the shape $\phi(x)$ we have to invert
\begin{eqnarray}
 x(\phi)= \int_{\phi} ^{\psi}\frac{d\tilde \phi}{\sqrt{{-2 \mathcal V(\tilde \phi)}}}  .
\end{eqnarray}
which follows by a quadrature from the pseudo-energy : $\frac{\phi'(x)^2}{2} + \mathcal {V}(\phi)=0$. The latter itself is derived from Poisson's equation.\\
 While (6) is the equation that determines the phase velocity $v_0$, it is (7) that delivers through (8) the wave structure $\phi(x)$, provided that the integral in (8) and the inversion can be accomplished by known mathematical functions otherwise one has to deal with a numerical evaluation of them. \\
An important feature of this system is that $ \Gamma $ (or $A$, respectively) no longer occurs in (7). This special trapping scenario therefore has no influence on the shape. Rather, it is this continuous variable that accounts for the phase velocity $v_0$ that accordingly belongs to a $\bf{continuous}$ dispersion relation.\\

It is easily seen that the parameter $k_0$ stands for periodic waves, namely either  from $\mathcal {V}(\phi)$ or directly from Poisson's equation. The curvature of $\phi$ : $\phi''(x)=n_e -n_i$, which becomes $\rightarrow k_0^2\psi$ as $\phi \rightarrow 0$,  vanishes in the solitary wave limit $k_0 \rightarrow 0$, noting that $\phi=0$ is the potential minimum. Otherwise the structure is periodic.
 It is, however, not necessarily the actual wave number $k$ which is defined by $k=\frac{\pi}{L}$ where 2L is the actual wavelength.
The correct relation between $k$ and $k_0$ is found by the overall charge neutrality condition and becomes to lowest order (\cite{DBS18,BDS18}):
\begin{eqnarray}
1=\frac{1}{2L}\int_{-L} ^{L} n_e dx=\frac{1}{L}\int_{0} ^{\psi} \frac{d\phi}{\sqrt{-2\mathcal {V}(\phi)}}
=\frac{k}{\pi}\int_{0} ^{1} \frac{d\varphi}{\mathcal {N}(k_0, B,...; \varphi) }
\end{eqnarray}
where $\mathcal {N}(k_0, B,...; \varphi) :=\frac{1}{\psi}\sqrt{-2 \mathcal {V}(\psi \varphi)}=
\sqrt{k_0^2\varphi (1-\varphi) + \varphi^2 [B(1-\sqrt\varphi) + ...]}$ and
where the dots stand for the remaining terms in $\mathcal {V}(\phi=\psi \varphi)$.\\

Equations (6)-(9) constitute our main result in its most general form.\\

We emphasize that $ \mathcal {V} (\phi) $ consists of 5 independent contributions, each of which stands for a certain mode structure. Whereas $k_0$ alone stands for the harmonic wave (or more generally in combination with the other parameters for periodic waves, as said), the other 4 themselves represent specific solitary waves.\\
It is this central functionality that gives us the opportunity to denote them by an own name: $ \bf {elementary} $ $ \bf {modes} $. With all 5 terms, however, when playing independently an active role in $\mathcal {V}(\phi)$, we have through all possible combinations a manifold of 31 ($\sum_{i=1}^{5}\frac{5!}{i! (5-i)!}$) wave modes of different provenance (5 single, 10 double, 10 triple, 5 quadruple and 1 quintuple combination(s)), a rather astonishing and up to now unknown variety. In principle the number of modes is doubled by the fact that besides $k_0=0$ there exists a second limit for $k_0$ which provides solitary modes (see Sect.IV.1 and Sect.V). \\
Unfortunately most of them are mathematically undisclosed because $ \phi (x) $ can no longer be found analytically. This typically holds when 3 or more combinations are involved, but also some of the double combinations suffer the same fate.
Fortunately, all of the elementary modes $ \phi (x) $ can be expressed and analyzed mathematically, an additional signature of their fundamental  role. We should however stress that the choice of elementary functions is not unique, as for example other non-perturbative trapping scenarios could be selected and added as well (\cite{S20a,S20b}).\\
In the context of the nonlinear world of structure formation, these elementary modes offer via these possible combinations a kind of $\bf{superposition}$ $\bf{principle}$  on the  $\mathcal {V}(\phi)$ level that enable us to get new members, somehow analogous to the superposition principle in linear wave theory. Or more precisely, within the class of potential structures $\phi(x)$ of given $\phi_{min}=0$ and $\phi_{max}=\psi$, the linear combination of two different pseudo-potentials $\mathcal {V}_1(\phi)$ and $\mathcal {V}_2(\phi)$ with corresponding $\phi_1(x)$ and $\phi_2(x)$ result in a third,  $\phi_3(x)$, which is provided by  $\mathcal {V}_3(\phi)=\mathcal {V}_1(\phi) +\mathcal {V}_2(\phi)$. 

\newpage

\section{The gallery of elementary modes}
III.1 The harmonic mode (single wave)\\

The harmonic,  monochromatic or single wave is obtained when $(B,\tilde C,D_1,D_2) \rightarrow 0$. In all five examples, however,  we will keep the $\Gamma$ trapping term which appears in the NDR only. We then have from (7), (8) and by inversion of (8)

\begin{eqnarray}
-\mathcal {V}(\phi)= \frac{k_0^2}{2} \phi(\psi-\phi)  \qquad   x(\phi)=\frac{1}{k_0}[\frac{\pi}{2} + \sin^{-1}(1-\frac{2\phi}{\psi})] \qquad \phi(x)=\frac{\psi}{2} [1+ \cos(k_0x)].
\end{eqnarray}

The surprising property of this mode is that it remains nonlinear up to the infinitesimal amplitude limit $ \psi \rightarrow 0^{+} $ (\cite{ SMS20b}). The reason is that as long as $ \psi \ne  0 $, there exists a non-vanishing trapping area of width $ 2 \sqrt {2 \phi} $ in phase space in which $ f_{et}$ behaves regularly. As seen from (1) $ f_{et}$ neither collapses to a $ \delta $-function (van Kampen) nor does it become a singular principal value function or  the more regularized perturbed function that is forced by an artificial and therefore unrealistic flattening of $ f_0 (v) $ at resonance (Landau). In our theory, in which these singularities or artificial interventions are obviously missing, the functional space is well posed. \\

Consequently, all linearly based single-wave models, as examined exemplarily in \cite {BMT13} as an extension of Landau or van Kampen, belong to this category of nonlinearly invalid models and must therefore be discarded.\\

The NDR (6)  becomes in this non-perturbative harmonic wave  limit: 
\begin{eqnarray}
k_0^2  -\frac{1}{2} Z_r'(\frac{ \tilde v_D}{\sqrt 2}) - \frac{\theta}{2} Z_r'(\frac{ u_0}{\sqrt 2}) = -\Gamma.
\end{eqnarray}.

Our mode is therefore the correctly upgraded, nonlinear counterpart to
Landau ($\Gamma=0$) and to van Kampen ($-\Gamma=\lambda$). It is moreover for $\Gamma=0$ the well-known "Thumb-Teardrop" DR which has been studied for $v_D=0$ in detail by \cite{TG18} mistakenly believing that it is a linear DR. As explained by the author in a comment in \cite{S19}, however, it makes sense only in the nonlinear regime, although formally it exists linearly, too.\\
On the other hand, a $ \Gamma \ne 0 $ provides a new parameter resulting in a $ \bf {continuous} $ $\bf{spectrum}$ of possible solutions, analogous but of course different  to the continuous linear spectrum of van Kampen.\\
The existence condition for the harmonic wave  is that all trapping parameters are zero (except $\Gamma$). This particularly means that $B \sim (1-\beta- \tilde v_D^2) e^{-\tilde v_D^2/2} =0$. This is satisfied either for large $|\tilde v_D|$ by the exp-function (Langmuir mode) or by $\beta=1-\tilde v_D^2$ in case of finite or small values of $|\tilde v_D|$. Since the latter is typically larger than unity the trapping parameter $\beta$ is a negative quantity corresponding to a hole in phase space. The assumption of a flat trapped region, $\beta=0$, as often anticipated in the literature (e.g. Landau-Lifshitz and related literature), is hence  generally inconsistent.\\
 We hence have to conclude that there is no linear analogon  of the harmonic wave that can account for the microscopic details. \\
The two worlds of linear and nonlinear Vlasov equilibria are disconnected with no connection (bridge) between them.
Or to give another comparison: It's like prospecting for gold; if you dig in the wrong valley, no matter how hard you try, you will never be successful.\\
The fact that they agree macroscopically (in shape and velocity for vanishing $\Gamma$) does not imply that they are also  identical microscopically.\\ 
As will be pointed out in detail later (Sect.VII.2) this mode is linearly marginally stable  \cite{S18} for all $v_D$ in strong contradiction to Landau's theory. There is no critical drift velocity  $v_D*$ which discriminates between damped and growing perturbations  of harmonic equilibria. All harmonic or single mode equilibria, due to their nonlinear character, turn out robust to linear perturbations and propagate undamped with repect to small linear perturbations independent of $v_D$. If there is any growth it must be due to the higher harmonic part of the spectrum, such as in cnoidal or solitary waves.\\
It is moreover easily seen that $k=k_0$ for this harmonic mode i.e. $k_0$ is already the exact wavenumber.\\

III.2 The privileged $\sech^4(x)$ - solitary mode\\

The four remaining modes are obtained by setting $k_0=0$. They are hence solitary in character. We get from (7),(8), by letting $(\tilde C, D_1,D_2) \rightarrow 0$  the following three expressions:\\
\begin{eqnarray}
-\mathcal {V}(\phi)=   B  \frac{\phi^2}{2} (1- \sqrt{\frac{\phi}{\psi}}), \qquad x(\phi)=\frac{4}{\sqrt B} \tanh^{-1}(\sqrt{1-\sqrt\frac{\phi}{\psi}}), \qquad\phi(x)=\psi \sech^{4}(\frac{\sqrt B}{4}x) \\
\nonumber
\end{eqnarray}
 where the last term represents the shape and  follows by inversion of $x(\phi)$. It must hold: $B>0$ which determines for given $\tilde v_D$ the parameter $\beta$.\\
This special shape has been known since the earliest times of structure formation \cite{Gur68,S72,S73,S79,S82,S86}.\\
In contrast to the next two solitary modes, which rest on a logarithmic trapping scenario, it stays existent in the low amplitude limit, representing a privilege for this mode.\\
The phase velocity $v_0$  is obtained by the NDR (6)
\begin{eqnarray}
 -\frac{1}{2} Z_r'(\frac{ \tilde v_D}{\sqrt 2}) - \frac{\theta}{2} Z_r'(\frac{ u_0}{\sqrt 2}) =  B -\Gamma,
\end{eqnarray}
which has depending on $B$ and $\Gamma$ a wide range of  particularly interesting solutions for  current-carrying plasmas, as shown next. The only condition is that $B>0$ whereas $\Gamma$ can carry either sign. The general solution   requires numerical means especially for the continuous branches which, due to $B$ and $\Gamma$, are  a bit more complex than the already complex Thumb-Teardrop DR.\\
We choose for demonstration two branches that are far apart.\\

(i) the slow electron acoustic wave branch (SEAW)\\
This branch is obtained by assuming $|\tilde v_D|\sim O(1) $  and $|B-\Gamma|<<1$ in which case all three terms in (13) are small. Making use of the Taylor expansion of the $Z_r'(x)$ function:   $-\frac{1}{2} Z_r'(\frac{ \tilde v_D}{\sqrt 2}) \sim \frac{1.307 -|\tilde v_D|}{1.307}$ and of  $\frac{1}{2} Z_r'(\frac{u_0}{\sqrt 2}) \sim \frac{\delta}{\theta v_0^2} << 1$ we get
                     $|\tilde v_D| = 1.307 ( 1 - B + \Gamma)$.
This mode is hence placed on both sides of the shifted Maxwellian at a distance of 1.307 and the phase velocity $v_0$ is given by    $\qquad v_0= v_D \pm 1.307 (1- B + \Gamma)$. 

This mode is acoustic-like and was termed $\bf{slow}$ $\bf{electron}$ $\bf{acoustic}$ $\bf{wave}$ (SEAW) in analogy to the slow ion acoustic wave (SIAW) ocurring in the ion case, where the notion "ion acoustic wave" (IAW) has already been taken for the known linear branch \cite{ S86}.  Hence the expression  "electron acoustic wave" for this mode, as used in the literature, is at least misleading. But it is also wrong because no linear electron acoustic wave exists, as long as one disregards anisotropic temperatures or other background deviations.\\

(ii) the ion acoustic wave branch (IAW)\\
In this case $|\tilde v_D| \sim O(\sqrt{\delta})<<1$ and $u_0 \sim \sqrt{\theta}>>1$, i.e. we assume $\theta>>1$.
The Taylor expansions yield $-\frac{1}{2} Z_r'(\frac{ \tilde v_D}{\sqrt 2}) = 1 - \tilde v_D^2$ and  $\frac{1}{2} Z_r'(\frac{u_0}{\sqrt 2}) \sim u_0^{-2}$ from which follows 
  $\qquad u_0=\sqrt{\theta} (1 + \frac{B-\Gamma}{2})$,
\newline
 which is the ion acoustic branch corrected by $B$ and $\Gamma$.\\

The evolution equation for which (12) is a stationary solution is of Schamel type and becomes (assuming $v_D=0$)  for the SEAW branch (see (17) of \cite{S20b})
\begin{eqnarray}
\phi_t +1.307 \biggl(1 + \Gamma - B\frac{15}{8}\sqrt\frac{\phi}{\psi}\biggr) \phi_x - 1.307 \phi_{xxx}=0
\end{eqnarray}
 and for the IAW branch (see(49) of \cite{S72} or (15) of \cite{S73})
\begin{eqnarray}
\phi_t + \biggl(1 + \Gamma + B\frac{15}{16}\sqrt\frac{\phi}{\psi}\biggr) \phi_x +\frac{1}{2} \phi_{xxx}=0
\end{eqnarray}
in which we renormalized t ($t \rightarrow \sqrt \delta t$) i.e. time is now normalized by the ion plasma frequency.
Both equations are of Schamel type and  make it possible to track evolutionary changes in the privileged solitary wave during its propagation, especially when overtaking processes or frontal collisions in case of several humps occur. \\

We moreover quote that with (13) the electron density gets the simpler form:
\begin{eqnarray}
n_e(\phi)=1+ \frac{\theta}{2} Z_r'(\frac{ u_0}{\sqrt 2})\phi +B(1 -\frac{5B}{4\sqrt \psi}\sqrt \phi)\phi  .
\end{eqnarray}
The density expressions for $n_e$ in (16)  and for $n_i$ in (3) make it easier to approach the measured structures already on the macroscopic level by considering the curvature of $n_{e,i}$ at potential maximum.\\
Since it holds $n_s'(x)=n_s'(\phi)\phi'(x)$ and $n_s''(x)=n_s''(\phi)\phi'^2(x) + n_s'(\phi)\phi''(x)$, $s={e,i}$, it follows that $n_s''(x=0)=n_s'(\psi)\phi''(0)$ which is true because of $\phi'(0)=0$. With $ \phi '' (0) <0 $ we see that the sign of $ n_s' (\psi) $ determines the curvature of $ n_s (0) $ in the center, $ s = {e, i} $.\\
For the ion density we get from (3): $n_i'(\psi)=\frac{\theta}{2}Z_r'(\frac{ u_0}{\sqrt 2})$ and for the electron density from (16): $n_e'(\psi)=\frac{\theta}{2}Z_r'(\frac{ u_0}{\sqrt 2}) - \frac{7}{8}B$ . When the SEH is propagating at ion acoustic speed, i.e. $\frac{1}{2}Z_r'(\frac{ u_0}{\sqrt 2})>0$, the curvature of $n_i(x) $ is unconditionally negative at x=0, whereas  $n_e(x=0)$ changes its sign from negative to positive when $B$ exceeds $B_c:=\frac{4\theta}{7}Z_r'(\frac{ u_0}{\sqrt 2})>0$.\\
The ion density is therefore bell-shaped in x under all circumstances, while the electron density gets a central depression when $ B$ exceeds  $ B_c $.\\

As an application we refer to the series of subcritical plasma simulations by Mandal, Schamel and Sharma \cite{SMS17,MSS18,MSS20,SMS20a,SMS20b}, in which the latter case was omnipresent in all cases considered. As an example we refer to Fig.1 of \cite{SMS20b} and the corresponding data:\\
 $\theta=10, \delta^{-1}=1836, v_D=0.01 < v_D*=0.053, c_s:=\sqrt \frac {T_e}{m_i}=3.16, u_0=4.74, v_0=0.035, M:=\frac{u_0}{c_s}=1.5, 
\newline
 \frac{1}{2}Z_r'(\frac{ u_0}{\sqrt 2})=0.52,   \psi=5.2 \times 10^{-5}, B_c=0.59  $.\\
The NDR (13) is satisfied for $B=0.48 +\Gamma$ which, due to the presence of a central depression in $n_e$, has to be larger than $B_c= 0.59$ from which we conclude that $\Gamma$ has to exceed 0.11: $\Gamma>0.11$.
In theses simulations the $\Gamma$ trapping scenario was automatically activated in all runs, which we hence can conclude already on the density i.e. on the macroscopic level.\\
The microscopic details still depend on the parameter $B$. For $B=1>B_c=0.59$ we get from the B formula:\\ $B:=\frac{16}{15} b(\beta,  \tilde v_D) \sqrt \psi$  with $b(\beta, \tilde v_D):= \frac{1}{\sqrt \pi} (1- \beta -  \tilde v_D^2)e^{- \tilde v_D^2/2}$ and by use of the above data the corresponding value for $\beta$: $\beta=-230$.
The electron distribution is therefore rather strongly depressed at resonance, a fact that has also be seen numerically, see e.g. Fig.5 of \cite{SMS20b}.\\
However, we should remind the reader that this is not evidence that the identification of the structure is unambiguous, as other trapping scenarios, or combinations thereof, may also be responsible for the settled structure \cite{SMS20a}.  \\

III.3 The Gaussian $e^{-x^2}$ - solitary mode\\

In this case $\Gamma$ and $D_1$ are the only non-vanishing parameters and we get \cite{SMS20a,SMS20b}

\begin{eqnarray}
-\mathcal {V}(\phi)=   D_1  \frac{\phi^2}{2} \ln\frac{\phi}{\psi}, \qquad x(\phi)=\frac{2}{\sqrt{-D_1}}\sqrt {\ln{\frac{\psi}{\phi}}}, \qquad\phi(x)=\psi e^{D_1x^2/4} \\
\nonumber
\end{eqnarray}

valid for $D_1<0$. 
This special solitary wave which has mainly be used by space plasma physicists to interprete their data is non-perturbative in nature. This implies that it has no zero-amplitude limit in contrast to the previous privileged $\sech^4(x)$ solitary electron hole (SEH) as seen by the nonlinear dispersion relation (NDR) which becomes:

\begin{eqnarray}
 -\frac{1}{2} Z_r'(\frac{ \tilde v_D}{\sqrt 2}) - \frac{\theta}{2} Z_r'(\frac{ u_0}{\sqrt 2}) =  D_1(0.887-\ln\psi) -\Gamma=:\hat B -\Gamma,
\end{eqnarray}

$B$ in (13) is therefore replaced by $\hat B:= -D_1(\ln \psi-0.887)$ in (18), which is a negative quantity for small $\psi$.
The discussion of the NDR is therefore pretty much the same as the previous one. The only difference is that the new $ \hat B $ is negative instead of positive which can however easily be compensated by $\Gamma$.\\

Again we can attribute a SEAW branch for which  $\qquad v_0= v_D \pm 1.307 (1- \hat B + \Gamma)$\\ 
and an IAW branch for which $\qquad u_0=\sqrt{\theta} (1 + \frac{\hat B-\Gamma}{2})$,\\
and discuss the role of $\Gamma$.\\
The evolution equation that relates to the SEAW branch is (see (17) of \cite{S20b})
\begin{eqnarray}
\phi_t +1.307 \biggl(1 + \Gamma  +D_1(2 + \ln \frac{\phi}{4})\biggr) \phi_x - 1.307 \phi_{xxx}=0
\end{eqnarray}
and similarly for the IAW branch. \\
Note that the competition between the two solitary structures, $ B $ and $ \hat B $, was discussed in \cite{SMS20a} to explain a numerically measured structure. The first indication of a logarithmic dependence of the trapped electron distribution in the case of a Gaussian SEH was given by \cite{S72b}. To distinguish it from other evolution equations we may call it logarithmic Schamel-type equation.\\

III.4 The Second Order Gaussian $e^{-\sinh^2(x)}$ - solitary mode\\

In this case $\Gamma$ and $D_2>0$ are non-zero and we get \cite{S20a}

\begin{eqnarray}
-\mathcal {V}(\phi)=   D_2  \frac{\phi^2}{2} \ln^2(\frac{\phi}{\psi}), \qquad x(\phi)=\frac{-2}{\sqrt{D_2}}
\ln \biggl(\frac{\sqrt r}{\sqrt {r-\ln \frac{\phi}{\psi}} +\sqrt{-\ln \frac{\phi}{\psi}}}\biggr), \qquad\phi(x)=\psi e^{-r\sinh^2{(\frac{\sqrt D_2x}{2}})} \\
\nonumber
\end{eqnarray}
where $r:=1.773-2\ln\psi$. This mode was first considered by the author in \cite{S20a}. With this solution we can compare a second, independent, non-perturbative trapping scenario with the usual Gaussian scenario. The effect is that $ x $ in the latter simply has to be replaced by $ \sinh (x) $ in order to arrive at the new structure. That's why we call it quasi-Gaussian. As extensively investigated in \cite{S20a}, it has essentially the same properties as the usual Gaussian SEH and can therefore explain an observation in the same way as the Gaussian.\\
The NDR becomes:

\begin{eqnarray}
 -\frac{1}{2} Z_r'(\frac{ \tilde v_D}{\sqrt 2}) - \frac{\theta}{2} Z_r'(\frac{ u_0}{\sqrt 2}) = \hat{ \hat{ B}} -\Gamma,
\end{eqnarray}
where $ \hat{ \hat{ B}}:= D_2(-1.326 +1.773 \ln \psi -\ln^2\psi)$. This variable now takes on the role of B in the NDR discussion, which we leave to the reader. It is clear again that a transition $ \psi \rightarrow 0 $ is impossible. Seeds of this type do not allow solutions with infinitesimal amplitudes.\\
The second-order logarithmic Schamel-type evolution  equation reads in this case for the SEAW branch
\begin{eqnarray}
\phi_t +1.307 \biggl(\hat{\mathcal A} + \hat r D_2 \ln \frac{\phi}{\psi} + D_2 \ln^2\frac{\phi}{\psi}\biggr) \phi_x - 1.307 \phi_{xxx}=0
\end{eqnarray}
where $\hat r= 1 + 2 \ln\frac{\psi}{4}$ and $\hat{\mathcal A}$ is an extension of the constant, $D_2$ independent term in the factor of $\phi_x$ in (19) inclusively $\Gamma$, to be derived by the reader.\\
In the next chapter we will show that the simultaneous presence of $D_1$ and $D_2$ belongs to the class of disclosed solutions $\phi(x)$, i.e. an explicit $\phi(x)$ can be presented for this pair of trapping scenarios.\\

III.5 The $\sech^2(x)$ soliton\\

In this final case all trapping terms are assumed zero except ($\Gamma,\tilde C$). We hence get:

\begin{eqnarray}
-\mathcal {V}(\phi)= q  \frac{\phi^2}{2} (1- \frac {\phi}{\psi}), \qquad x(\phi)= \frac{2}{\sqrt{q}} \tanh^{-1} \sqrt{ 1-\frac{\phi}{\psi}}, \qquad \phi(x)=\psi \sech^2(\frac{\sqrt{q} x}{2}) 
\end{eqnarray}
where $q:=-\tilde C\psi$ and a solution exists  as long as $q>0$.\\
The corresponding NDR reads:

\begin{eqnarray}
 -\frac{1}{2} Z_r'(\frac{ \tilde v_D}{\sqrt 2}) - \frac{\theta}{2} Z_r'(\frac{ u_0}{\sqrt 2}) = q -\Gamma.
\end{eqnarray}
The discussion of the NDR therefore proceeds as in III.2 including the two branches SEAW and IAW. We just need to replace B with q, both of which must be positive. Of particular interest is the case of no $ \zeta $- trapping scenario (C = 0) for which we get:  $q=\frac{1}{24}\bigg(\theta^2 Z_r'''(\frac{u_0}{\sqrt 2})-Z_r'''(\frac{\tilde v_D}{\sqrt 2})\bigg)\psi$.\\
For the SEAW branch, $|\tilde v_D|=O(1), u_0>>\sqrt \theta$, we then get $q=\bigg(\frac{\theta^2}{u_0^4} +\frac{1}{6}\bigg)\psi\approx \frac{\psi}{6} >0$ which means that a positive q is automatically satisfied.\\

For the IAW branch, when it holds $|\tilde v_D|=O(\sqrt \delta), u_0 \sim \sqrt \theta$ and $v_D=0$, we have $q=(\frac{\delta^2}{v_0^4} -\frac{1}{3})\psi =2\psi/3>0$ which is positive either.  The common ion acoustic soliton is hence represented by (23). In this case $n_e=1 + \phi + \phi^2/2 +..$ and $n_i=1 + \phi +3\phi^2/2 +...$ and we have a complete match with the macroscopic fluid result which is thus recovered within the limits taken. 
The evolution equation in the IAW case, for which (23) is a solution, is given by
\begin{eqnarray}
\phi_t + \biggl(1 + \Gamma + \phi \biggr) \phi_x +\frac{1}{2} \phi_{xxx}=0,
\end{eqnarray}
where again time is renormalized by the ion plasma frequency (i.e. $t \rightarrow \sqrt \delta t$).
This is (for $\Gamma=0$)  the well-known, integrable Korteweg de Vries equation. \\

We however stress that microscopically we have an abundance of $\sech^2$- solutions  belonging to the continuous spectrum not only because $\Gamma$ may be nonzero but also because of the various additional continuous solutions to the NDR that supplement the SEAW and IAW analytical approach.  Moreover, since $q=-\tilde C \psi=-\psi \bigg( \frac{2C}{3} +\frac{1}{24}[ Z_r'''(\frac{ \tilde v_D}{\sqrt 2}) -\theta^2 Z_r'''(\frac{ u_0}{\sqrt 2})]   \bigg)$
 there is through the $Z_r'''$ terms always a nonzero contribution to $\mathcal V(\phi)$ which stems from the free electron and ion  distributions, respectively, even when $C=2 \zeta/3$ is negligible. Therefore even if all trapping scenarios are   negligible ($\gamma= \beta= \chi_1=\chi_2= \zeta=0$) we still have a finite trapped electron region of width $2\sqrt 2\psi$ where $f_{et} \sim \{ 1+...\}$ is nonzero and the $sech^2$- solution  keeps his microscopic nature.  In VP plasmas inhomogeneous equilibria are intrinsically nonlinear and of course microscopic. The embedding of the $\sech^2$- fluid solution in the continuous spectrum has to be seen this way, namely as a special microscopic solution. A proof of its existence can hence only be given kinetically. \\

This section was devoted to isolated single trapping scenarios yielding to what we called elementary modes. As said, by combinations new solutions can be obtained. In the next section examples are presented in which two trapping scenarios are in action at the same time and which lead to new patterns through suitable combinations. As before, $ \Gamma $ is treated independently, since it has disappeared in $ \mathcal V (\phi) $. Three of the possible combinations will have disclosed potentials $ \phi (x) $, whereas one will appear with an undisclosed $ \phi (x) $. \\

\section{ Holes caused by two trapping scenarios}

IV.1 The cnoidal electron hole (CEH) and the solitary hole (SEH) of negative polarity\\

Periodic EH solutions are obtained by non-zero ($k_0, B$) in (7) with vanishing ($D_1, D_2, \tilde C)$. The trapping scenario $\Gamma$ in $A$ is retained in order to obtain maximum variability of the possible phase velocities. We then have from (7):\\
\begin{eqnarray}
-\mathcal {V}(\phi)= \frac{k_0^2}{2} \phi(\psi-\phi) +   B  \frac{\phi^2}{2} (1- \sqrt{\frac{\phi}{\psi}})  
\end{eqnarray}
As shown in \cite{KS96a}, equations (3.24)-(3.29), there exist three different regions in which $\phi(x)$ is represented by Jacobian elliptic functions. They are distinguished by the parameter $\hat L:=\frac{k_0^2}{4B}$ and are given by :
$\hat L < -\frac{1}{8}$, $0 \le \hat L \le 1$, and $1 < \hat L $. This implies that negative $Bs$ are now admitted.\\
Another characterization can be made by the "steepening" parameter $S:=\hat L^{-1}$ ,   \cite{S72}, which variies between -8 and infinity : $-8 \le S \le \infty$. From $n_e$ it is seen that one has rarefactive waves when $S \ge 0$ and compressional waves when $S<0$. $S\rightarrow 0$ yields the harmonic wave (10) (no steepening!), and  $S\rightarrow \infty$ results in the hump-shaped solitary EH (12) (maximum steepening!).\\

 Of particular interest is the lower limit of $ S $: $S=-8$ or $B=-2k_0^2$, in which case (26) becomes\\
$-\mathcal {V}(\phi)= \frac{k_0^2}{2\sqrt \psi}( \phi \psi^{3/2} -3\phi^2  \psi ^{1/2}+ 2 \phi^{5/2}) =\frac{k_0^2 \psi^2}{2} \varphi (1-\sqrt \varphi)(1+\sqrt \varphi -2\varphi) $ where $\varphi:=\phi/\psi$. The last expression shows that $\mathcal {V}$ has a double zero at $\varphi=1$.\\

The corresponding $\phi(x)$  becomes for $x \ge 0$:

$\phi(x)= \psi \bigg [\frac{2\sinh\zeta(2\sinh\zeta +\sqrt 3\cosh\zeta)}{(\sinh\zeta + \sqrt3 \cosh\zeta}\bigg]^2$
where $\zeta:=\frac{\sqrt3 k_0x}{4}$. It is given by (3.33) in \cite{KS96a} and is rederived in Appendix B, in which we offer an expression that is valid for arbitrary x:
\begin{eqnarray}
\varphi(\zeta)= \frac{1}{4}\bigg[ 3 \tanh^2(|\zeta| + \zeta_0) -1 \bigg]^2
\end{eqnarray}
where  $\zeta_0:= \tanh^{-1} (\frac{1}{\sqrt{3}})=0.65848  $.  \\ 

\begin{figure}
\includegraphics[width =  0.6 \textwidth] {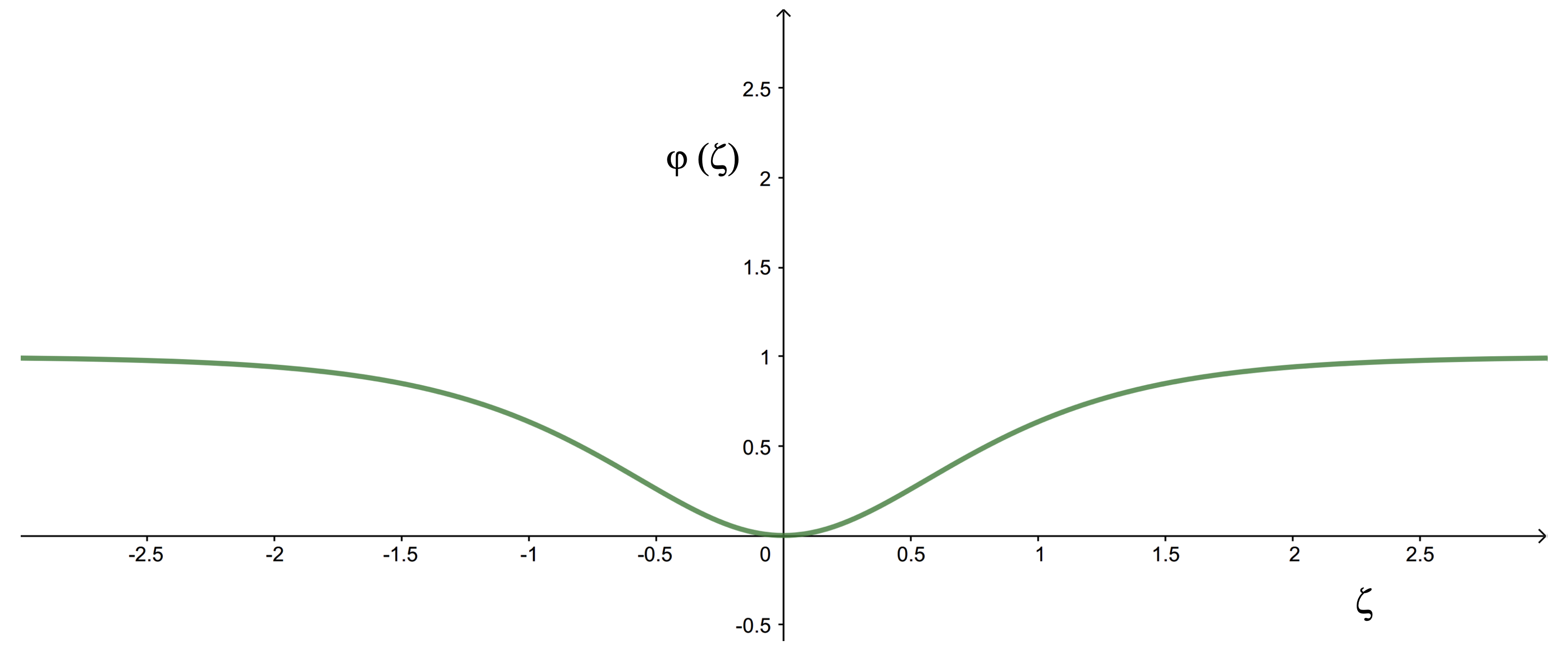}
\caption{Solitary electron hole of negative polarity $\varphi(\zeta)$ as a function of $\zeta$}
\end{figure}

  Fig.1, in which   $\phi/\psi$  is plotted in the interval $-3\le \zeta \le +3$, shows that it has an unexpected negative polarity like a solitary ion hole \cite{SB80,BS81}.\\

We mention that in a recent statistical analysis of more than two thousand "bipolar electrostatic solitary waves" (ESW) \cite{Wang21} collected from ten quasi-perpendicular Earth's bow crossings, about 95$\%$ of the ESWs were found of $\bf{negative}$ polarity. Since the phase velocities were in the order of the local ion-acoustic velocity, the authors argued that these must have been solitary ion holes \cite{SB80, BS81}. This determination is too premature, however, as SEHs of negative parity can also come into question, as presented in this section. This interpretation is possibly as relevant as the ion hole interpretation since linear ion-ion streaming instabilities  are not necessarily required for both.  We remind the reader that Landau theory must not necessarily hold for holes growing out of seeds, but of course larger drifts facilitate their excitation.\\

Altogether, there is therefore an abundance of cnoidal hole solutions that are characterized by a single parameter $ S $ and that become solitary-like at both borders with opposite polarity.\\
The phase velocity $v_0(u_0)$ follows from the the NDR (6), which becomes
\begin{eqnarray}
 -\frac{1}{2} Z_r'(\frac{ \tilde v_D}{\sqrt 2}) - \frac{\theta}{2} Z_r'(\frac{ u_0}{\sqrt 2}) = B -\Gamma- k_0^2 \sim -3k_0^2 - \Gamma
\end{eqnarray}
It is therefore of the same type as the previously discussed cases and the continuous spectrum is again controlled and expanded by the additional trapping parameter $ \Gamma $.\\
The last step in (28) applies to $ S = -8 $, to which we now turn our attention. We provide the corresponding densities and phase velocities for the two analytic branches SEAW and IAW.\\

 For SEAW, when to lowest order $|\tilde v_D|\sim 1.307$ and $u_0\sim \sqrt{\theta/\delta}$, we have $n_e=1+k_0^2(\psi/2 -3\phi +\frac {5}{2}\phi\sqrt{\phi/\psi})$ and $n_i \approx 1$, whereas in the IAW case, when $|\tilde v_D|\sim O(\sqrt \delta)$ and $u_0\sim \sqrt{\theta}$, we have  $n_e=1+k_0^2(\psi/2 -3\phi +\frac {5}{2}\phi\sqrt{\phi/\psi}) + \phi$ and $n_i=1 +\phi$. In both cases $n_e$ is hump-like and the electrons experience compression, whereas the ion densities are either nearly constant for SEAW (a too fast phase velocity for the ions to react!) or dip-like for IAW (note that $\phi=\psi$ is then at infinity!).\\

In order to obtain the corresponding evolution equation, for which $ \phi (x-v_0t) $ is a solution, we use the method proposed in \cite{S20b} by "adding two zeros" via a coupling constant $\mathbf{c}$ :
$[\phi_t +v_0 \phi_x] + \mathbf{c} [-\mathcal V''(\phi) \phi_x - \phi_{xxx}]=0$.\\ 
 We only treat a current-less plasma $v_D=0$.\\

For SEAW  we get with $v_0=1.307(1 + \Gamma +3k_0^2)$ and $\mathbf{c}=1.307$ the Schamel-type evolution equation for this special solitary EH of negative polarity: 
\begin{eqnarray}
\phi_t + 1.307 \biggl(1 + \Gamma + \frac{15 k_0^2}{4} \sqrt{\frac{\phi}{\psi}} \biggr) \phi_x  -1.307 \phi_{xxx}=0,
\end{eqnarray}
which is nearly identical with (14) despite the different physical background.\\

For IAW we get with $v_0=\sqrt \delta( 1- \frac{\Gamma+3k_0^2}{2})$ and  $\mathbf{c}= -\frac{\sqrt \delta}{2}$
 the following  Schamel evolution equation:
\begin{eqnarray}
\phi_t + \sqrt\delta \biggl(1 - \frac{\Gamma}{2} - \frac{15 k_0^2}{8} \sqrt{\frac{\phi}{\psi}} \biggr) \phi_x +\frac{\sqrt \delta}{2} \phi_{xxx}=0,
\end{eqnarray}
which is equivalent to (15) if we again renormalize time: $t \rightarrow \sqrt \delta t$. Note that in both cases the different polarity of $ \phi (x) $ is reflected in the sign of the nonlinear term. \\
In both regimes, the corresponding trapping parameter $\beta$ is a function of $k_0^2$ and $\psi$ and follows from $-2k_0^2= B =\frac{16}{15} b(\beta,  v_0) \sqrt \psi$  with $b(\beta,  v_0):= \frac{1}{\sqrt \pi} (1- \beta -   v_0^2)e^{-  v_0^2/2}$.\\

We note that if $ B $ is replaced by one of $ D_1, D_2 $ or $ \tilde C $, three new series of periodic hole solutions with solitary wave character at the two boundaries for each parameter are obtained. \\
In Sect. V we shall briefly address the whole class solitary electron holes of negative polarity.\\

IV.2 The  Schamel-Korteweg de Vries solitary electron hole (SKdV-SEH) \\

For the next two-parametric solution we choose $ B $ and $ q = - \tilde C \psi \equiv \frac{2\psi}{3}$ in (7) as the only non-zero parameters and obtain
\begin{eqnarray}
-\mathcal {V}(\phi)=     \frac{\phi^2}{2}\biggl[   B (1- \sqrt{\frac{\phi}{\psi}})  +  q (1-\frac{\phi}{\psi}) \biggr].
\end{eqnarray}

The corresponding NDR reads
\begin{eqnarray}
 -\frac{1}{2} Z_r'(\frac{ \tilde v_D}{\sqrt 2}) - \frac{\theta}{2} Z_r'(\frac{ u_0}{\sqrt 2}) =  B - \Gamma -q.
\end{eqnarray}
This case has already been treated as early 1972 in \cite{S72}, equations (47), (48), with the result that $\phi(x)$ is given by
\begin{eqnarray}
\phi(x)= \psi \frac{\sech^4(y)}{[1+\frac{\tanh^2(y)}{1+B/q}]^2}
\end{eqnarray}
where $y:=\frac{x}{2}\sqrt{\frac{\psi}{6}(1 + \frac{B}{q})}$ and $-q < B$.\\
For $1<<\frac{B}{q}$ and $|\frac{B}{q}|<<1$, respectively, this expression reduces  to the known cases (12) and (23), respectively, the privileged $\sech^4(x)$ and the KdV solitary wave. The NDR (32) can be discussed like the previous cases.\\
In the IAW limit, the following evolution equation, which has (33) as the equilibrium solution, can be easily derived ($t \rightarrow \sqrt \delta t)$:
\begin{eqnarray}
\phi_t + \biggl(1 + \Gamma + \frac{5\sqrt3 B}{8\sqrt2 \sqrt q} \sqrt{\phi}  + \phi \biggr )\phi_x  -\frac{1}{2} \phi_{xxx}=0,
\end{eqnarray}
which is (49) of \cite{S72}. This Schamel-Korteweg de Vries equation reduces in the appropriate limits to (15) and (25), respectively, as expected.\\

IV.3 The modified second order Gaussian SEH \\

In this part we refer to the Gaussian SEH in its first and second order version and use

\begin{eqnarray}
-\mathcal {V}(\phi)=   \frac{\phi^2}{2}\biggl[(D_1-r D_2) \ln(\frac{\phi}{\psi}) +D_2 \ln^2(\frac{\phi}{\psi})\biggr]
\end{eqnarray}
where $r:=1.773-2\ln\psi$ \cite{S20a}, and get  for $x(\phi)$ with $s:=r-\frac{D_1}{D_2}>0$
\begin{eqnarray}
\sqrt D_2  x(\phi)= -2 \ln\biggl(\frac{\sqrt s}{\sqrt{s-\ln\frac{\phi}{\psi}} +\sqrt{-\ln\frac{\phi}{\psi}}} \biggr).
\end{eqnarray}
Its inversion yields
\begin{eqnarray}
\phi(x)=\psi e^{-s \sinh^2{(\frac{\sqrt D_2x}{2}})} .
\end{eqnarray}
It reduces to the ordinary Gaussian SEH (17) in the limit $D_2 \rightarrow 0$  ($sD_2 \rightarrow -D_1$, resp.), and to the second order Gaussian SEH (20) in the limit $D_1 \rightarrow 0$.\\
The NDR (6) is simplified in this case of $k_0=B=\tilde C=0$ and becomes
\begin{eqnarray}
 -\frac{1}{2} Z_r'(\frac{ \tilde v_D}{\sqrt 2}) - \frac{\theta}{2} Z_r'(\frac{ u_0}{\sqrt 2}) = - \Gamma +\hat{ B} + \hat{ \hat{ B}},
\end{eqnarray}
where $\hat{B} $ and  $\hat{\hat{B}} $ had been defined in connection with equations (18) and (21), respectively.\\
The evolution equation for the SEAW branch becomes
\begin{eqnarray}
\phi_t +1.307 \biggl(1 + \Gamma +\hat{\hat{\mathcal A}} + (D_1+\hat r D_2) \ln \frac{\phi}{\psi} + D_2 \ln^2\frac{\phi}{\psi}\biggr) \phi_x - 1.307 \phi_{xxx}=0,
\end{eqnarray}
where $\hat{\hat{\mathcal A}}:=\hat{\mathcal A} + D_1(2 + \ln{\frac{\psi}{4}})$. \\

We note that in the three cases treated so far we were able to solve the decisive equation (8) in order to obtain $x(\phi)$  by inversion $\phi (x) $, i.e. our solution $ \phi (x)$ could be expressed by known mathematical functions. It was disclosed. In our final example, we will encounter a situation where this disclosure no longer exists. It represents the general case. \\

IV.4 The undisclosed logarithmic Schamel  SEH \\

This is the case when the two basic trapping scenarios $ (B, D_1)$ and only the two  are in action simultaneously. The pseudo-potential $\mathcal V(\phi)$ then reads
\begin{eqnarray}
-\mathcal {V}(\phi)=  \frac{ \phi^2}{2} \biggl(  B (1- \sqrt{\frac{\phi}{\psi}}) + D_1 \ln(\frac{\phi}{\psi})\biggr),
\end{eqnarray}
a case which has been treated thoroughly in \cite{S20b}. The integral for $x(\phi)$, given by (9) in \cite{S20b}, cannot be solved anymore. We can conclude from this that these two trapping channels lead us into an unknown terrain, into an area where only  numerically  an image of the potential $\phi(x)$ can be obtained.\\
This is fortunately not the case for the NDR and the evolution equation, which become

\begin{eqnarray}
 -\frac{1}{2} Z_r'(\frac{ \tilde v_D}{\sqrt 2}) - \frac{\theta}{2} Z_r'(\frac{ u_0}{\sqrt 2}) = B - \Gamma - D_1[\frac{1}{2}-2\ln 2 +\ln \psi],
\end{eqnarray}
and
\begin{eqnarray}
 \phi_t  +1.307\biggl [\mathcal A - B\frac{15}{8}\sqrt{\frac{\phi}{\psi}} + D_1 \ln {\frac{\phi}{\psi}}\biggr]\phi_x - 1.307 \phi_{xxx}=0
\end{eqnarray}
with $\mathcal A $  given by
\begin{eqnarray}
 \mathcal A= 1 + A + D_1(1 +  \ln \psi) =1 + \Gamma + D_1(2 + \ln{\frac{\psi}{4}})
\end{eqnarray}
the latter being valid for the SEAW branch. It may be named for clear identification logarithmic Schamel equation. It is therefore worth noting that although the explicit form of $ \phi (x) $ is not known, an evolution equation can still be assigned.\\

We conclude that a large number of new structures are already coming into play for two trapping channels in action. This is all the more true when more than two scenarios are involved. This abundance of electrostatic structures is a consequence of the nonlinear treatment of the Vlasov equation (s) with no chance of a linear approximation.\\
In the following two special cases are discussed  before their setting in a more general context will be discussed.\\

\section{ The class of negatively polarized solitary electron holes (SEHs)}

In Sect. IV.1 we learned that by setting the two parameters $ B $ and $ k_0 ^ 2 $ adequately, SEHs with negative polarity can be obtained. Motivated by its ubiquity in extraterrestrial space \cite{Wang21} we extend this two-parametric solution to all five parameters to get the general class of SEHs with negative polarity. The condition for the five parameters $ (k_0, B, D_1, D_2, q) $ to achieve a negatively polarized SEH is obtained by setting $ \mathcal {V} '(\psi) = 0 $, such  that the double zero point goes from $\phi=0$ over to $ \phi = \psi $.\\
This constraint follows from (4) in which we replace the first  bracket by (6) and by setting $ \mathcal {V} '(\psi) = 0 $ to get

\begin{eqnarray}
k_0^2= -\frac{B}{2} + D_1 + (a_1-1+\ln\psi) D_2 - q
\end{eqnarray}

where $q:=-\tilde C \psi$. Replacing $k_0^2$ in (7) for  $ \mathcal {V} (\phi)$  by (31) we get

\begin{eqnarray}\nonumber
- \mathcal {V} (\phi)/\psi^2= \frac{-B}{4} \varphi (1-3\varphi +2 \varphi^{3/2}) + \frac{D_1}{2} \varphi (1-\varphi +\varphi \ln \varphi)\\ \nonumber+ \frac{D_2}{2} \varphi \bigg [ (a_1 - 1+\ln \psi)(1-\varphi) + \varphi \ln^2\varphi)\bigg ]\\  -\frac{q}{2} \varphi (1-\varphi)^2.
\end{eqnarray}
By replacing $k_0^2$ in (6) through (31) we get the associated NDR 

\begin{eqnarray}
\biggl( A -\frac{1}{2} Z_r'(\frac{ \tilde v_D}{\sqrt 2}) - \frac{\theta}{2} Z_r'(\frac{ u_0}{\sqrt 2}) \biggr ) - \frac{3B}{2} +D_1 (\frac{1}{2} + \ln\psi) + D_2\bigg [\frac{a_1-1}{2} + a_1\ln \psi) -\ln^2\psi \biggr] -2q=0.
\end{eqnarray}

With the two equations (45) and (46) we have all the ingredients for a complete theory of SEHs with negative polarity.
With four independent parameters $(B, D_1, D_2, q)$ we have a 15-fold manifold of different solutions ($\sum_{i=1}^{4}\frac{4!}{i! (4-i)!}=15$). Most of them appear as mathematically undisclosed solutions.\\
Only two special cases are treated further.\\
One simple case is when only $q$ is present. For $B=D_1=D_2=0$ we then have $ \mathcal {V} (\phi)=\frac{q\psi^2}{2} \varphi (1-\varphi)^2$. Since it must be negative we have $q=-\tilde C \psi<0$ or $\tilde C>0$ as a requirement for the existence of a solution. The potential is given by $\varphi(x)= 1- \sech^2(\frac{\sqrt{-q}x}{2})$, an expected result (see equation (23) where $\varphi$ has to replaced by $1-\varphi$ to yield the new result).\\
The other simple case is that only $D_1$ is present, in which case we have $- \mathcal {V} (\phi)/\psi^2= \frac{D_1}{2}
 \varphi (1-\varphi +\varphi \ln \varphi)$. Insertion  into  $x(\phi)=\pm \int_{0} ^{\phi}\frac{dt}{\sqrt{{-2 \mathcal V(t)}}}$ (see Appendix B) yields  $\sqrt D_1 x=  \pm 2 \int_{0} ^{\varphi} \frac{dy}{\sqrt{ 1 -  y^2 + 2 y^2 \ln y}}$, $D_1>0$. In contrast to Sect. III.3, the Gaussian solitary mode, where $D_1<0$, $D_1$ must now be positive for a solution to exist. However, we have not been able to solve this latter integral. So it seems that this particular negatively polarized SEH that definitely exists is already part of the plethora of undisclosed potential patterns like most others are.\\
 We won't pursue any further details, leaving it up to the reader to use them in any particular case.\\

We finally note that this class was obtained by imposing the additional constraint $ \mathcal {V} '(\psi) = 0 $. By setting further restrictions, new structures with different characteristic shapes can be created. An example is the additional constraint $ \mathcal {V} '(0) = 0 $ which leads to well-known double layer (DL) \cite{SB83,S86}.
The more trapping parameters are involved, the more constraints  can be imposed. It would therefore be of interest to what kind of structure the additional condition   $ \mathcal {V}''(0) = 0 $ and/or  $ \mathcal {V}''(\psi) = 0 $ is leading.
Future scientific generations may take up this problem.\\

\section{ The class of ultra slow SEHs}

For a given occasion, we would like to draw our attention to another issue, that of the extremely slow SEHs. In a recently published article \cite{Hu21} the opinion was spread that the current theory has flaws which is particularly evident in the lack of ultra-slow SEHs for single humped ion distributions. Here, equipped with the correct method, we show the opposite, namely the existence of ultra-slow SEHs for a single-humped $ f_i $.\\
For the sake of simplicity, in the following we assume an ordinary SEH, namely one with a positive hump and  take $k_0^2=0$.\\
The NDR in (6) therefore becomes
$ \biggl( A -\frac{1}{2} Z_r'(\frac{ \tilde v_D}{\sqrt 2}) - \frac{\theta}{2} Z_r'(\frac{ u_0}{\sqrt 2}) \biggr ) - B + \hat D =0$ where $\hat D:=[D_1 + (a_1-1)D_2](-\frac{1}{2} +\ln \psi) + D_2\ln^2\psi + \tilde C \psi \biggr]$
 , which in case of a non-propagating structure: $v_0=0=u_0$ simplifies to 
\begin{eqnarray}
 \biggl( A -\frac{1}{2} Z_r'(\frac{  v_D}{\sqrt 2}) +\theta \biggr ) = B - \hat D .
\end{eqnarray}
 We remember that $A$ was defined by $A:=(\Gamma +\frac{a_1}{2}D_1 +a_2D_2)$. With given $ v_D $ and $ \theta $ this equation represents the condition which the remaining parameters $ (\Gamma, B, D_1, D_2, \tilde C$ (or $q$))  have to fulfill. This corresponding 5-parameter solution set again has 31 members and is therefore not insignificant. So there can be no question of a missing solution. Especially due to the presence of $ \Gamma $, a solution can always be found. In case of a vanishing drift $v_D=0$ and of vanishing parameters $(\Gamma, D_1, D_2, q)$ except $B$ we have
$1+\theta = B =\frac{16}{15}\frac{1}{\sqrt \pi} (1- \beta)$ which holds for any $\theta$ inclusively $\theta = \frac{T_e}{T_i}<1$. Hence a $\beta=-(0.66 +1.66 \frac{T_e}{T_i} )$ and with it a sufficiently excavated trapped electron distribution will do the job. We don't need ion trapping effects to get a solution. (In parenthesis we state that this does not mean the absence of ion trapping, only the absence of its effects, ($\alpha=1$ see later)). In a recently conducted VP simulation, such structures could undoubtedly be demonstrated \cite{MSS20}.\\
 If ion trapping/reflection effects come into play the NDR is modified and we have to look at altered solutions. But there is no doubt that solutions do exist as well.\\
The continuous spectrum is extremely rich in elements and provides holes with almost arbitrary phase velocities, which are sustained by appropriately adapted trapping scenarios.\\

\section{ Ion trapping effects and ion holes}

VII.1 Ion trapping effects\\

In this section we briefly discuss the effects of ion trapping, which can be important, for example, for the slow propagation of electron holes in the ion thermal range. In addition, we will briefly address the existence of ion holes.
The incorporation of ion trapping (reflection) effects can be straightforwardly performed by the following replacements in  (1) to get $f_i(x,u)$: namely  $\varepsilon:=\frac{v^2}{2}-\phi$ by  $\epsilon:=\frac{u^2}{2}-\theta( \psi-\phi)$  ; $\tilde v_D$ by $u_0$ ;  a change in the normalization $1+ \frac{k_0^2 \psi}{2}$ by $1+ K_i$ and by attaching an index i to the new ion trapping parameters: $\Gamma_i, B_i, C_i, D_{1i},D_{2i}$. Note that $\beta$ becomes $\alpha$.
The details of this procedure are found  in \cite{S00}, especially in Sect.IV of this paper. We then have:

\begin{eqnarray}\nonumber
  f_i (x,u)=\frac{1+ K_i}{\sqrt{2\pi}}  \biggl(\theta(\epsilon) \exp [- \frac{1}{2}(\sigma \sqrt{2\epsilon} - u_0)^2] + \\
\theta(-\epsilon) \exp(-\frac{ u_0^2}{2}) \{1 + [\gamma_i +\chi_{1i} \ln(-\epsilon) +\chi_{2i} \ln(-\epsilon)^2] (-\epsilon)^{1/2} -\alpha \epsilon +\zeta_i (-\epsilon)^{3/2} \} \biggr).
\end{eqnarray}
The corresponding density  then becomes
\begin{eqnarray}\nonumber
n_i(\phi)=(1 + K_i)\biggl[1+\biggl( A_i
  -\frac{1}{2} Z_r'(\frac{u_0}{\sqrt 2})  -\frac{5B_i}{4\sqrt \psi}\sqrt{\theta(\psi- \phi)}  + C_i \theta(\psi- \phi) + (D_{1i} +a_1D_{2i})  \ln{\theta(\psi- \phi)}   \\
+ D_{2i}  \ln^2{\theta(\psi- \phi)}  \biggr)\theta(\psi- \phi)  + \frac{1}{16} Z_r'''(\frac{u_0}{\sqrt 2}) \theta^2(\psi- \phi)^2 \biggr],
\end{eqnarray}

where we defined: $A_i:=\Gamma_i + \frac{a_1}{2}D_{1i} + a_2 D_{2i}$ , $B_i:=\frac{16}{15} b(\alpha, u_0)\sqrt \psi$ with $b(\alpha, u_0):=\frac{1}{\sqrt \pi}( 1- \alpha- u_0^2)e^{-u_0^2/2}$  and $(\Gamma_i, C_i, D_{1i}, D_{2i}):= \frac{\sqrt \pi}{2} e^{-u_0^2/2} \biggl( \gamma_i, \frac{3\zeta_i}{4}, \chi_{1i}, \chi_{2i} \biggr)$.\\
The normalization constant $ K_i $ is determined by the requirement that in the solitary wave limit $k_0 \rightarrow 0$
both densities $n_e$ and $n_i$ should be equal (namely unity) at infinity when $\phi \rightarrow 0$, which yields:

\begin{eqnarray}\nonumber
1=(1 + K_i)\biggl[1+\biggl( A_i
  -\frac{1}{2} Z_r'(\frac{u_0}{\sqrt 2})  -\frac{5B_i}{4\sqrt \psi}\sqrt{\theta\psi}  + C_i \theta\psi + (D_{1i} +a_1D_{2i})  \ln{\theta\psi}   \\
+ D_{2i}  \ln^2{\theta\psi}  \biggr)\theta\psi  + \frac{1}{16} Z_r'''(\frac{u_0}{\sqrt 2}) \theta^2\psi^2 \biggr].
\end{eqnarray}
To check this expression we take the zero limit of $(A_i, D_{1i},D_{2i}, Z_r'''(\frac{u_0}{\sqrt 2})$ and get
$1=(1 + K_i)\biggl[1+\biggl( 
  -\frac{1}{2} Z_r'(\frac{u_0}{\sqrt 2})  -\frac{5B_i}{4\sqrt \psi}\sqrt{\theta\psi} 
  \biggr)\theta\psi   \biggr]$ from which follows $K_i=\biggl( 
  \frac{1}{2} Z_r'(\frac{u_0}{\sqrt 2})  +\frac{5B_i}{4\sqrt \psi}\sqrt{\theta\psi} 
  \biggr)\theta\psi  $ which is identical with (21) of \cite{S00}.\\

Replacing  $K_i$ in  $n_i$ by this general expression obtained from (50), we can proceed as before: we determine through $\phi''(x)=n_e - n_i = -\mathcal V'(\phi)$ the preliminary form of $ V(\phi)$: $ V_0(\phi)$ (such as in (5)) to get through
 $ V_0(\psi)=0$ the NDR analogue of (6), we may call (6'). Removing finally the bracket in  $ V(\phi)$ which involves $(v_0,u_0)$ through (6'), we can finally find the desired expression for $ V(\phi)$, called (7'), in which both the electron and ion trapping effects are incorporated on equal footing.\\
We will neither write it down [but may call it (7 ')] nor deduce the general consequences, leaving this interesting and straightforward but cumbersome procedure to the reader (and perhaps later generations).\\
Without ionic trapping effects we had 1+4=5  individual terms in $ V(\phi)$ corresponding to $\sum_{i=1}^{5}\frac{5!}{i! (5-i)!}=31$ possible combinations and hence 31 patterns that can be distinguished. With ion trapping effects the number of free trapping parameters is enhanced by further 4 such that $\sum_{i=1}^{9}\frac{9!}{i! (9-i)!}=494$ individual modes 
become vivid, without taking into account the double counting of the solitary waves of positive and negative polarity.\\

What we finally want to show in this section is the existence of ion holes of positive polarity.\\

VII.2 Ion holes of negative and positive polarity\\

In the limit of vanishing parameters $(A_s, C_s, D_{1s},D_{2s})$, s=e,i,  where the index e refers to the previous electron parameters, and of  $Z'''_r(\tilde v_D/\sqrt2) \simeq 0 \simeq Z'''_r(u_0/\sqrt2)$, the governing equations simplify to 
\begin{eqnarray}
k_0^2  -\frac{1}{2} Z_r'(\frac{ \tilde v_D}{\sqrt 2}) - \frac{\theta}{2} Z_r'(\frac{ u_0}{\sqrt 2}) = B_e + \frac{3}{2}\theta^{3/2} B_i
\end{eqnarray} 
and 
\begin{eqnarray}
-\mathcal {V}(\phi)/ \psi^2= \frac{k_0^2}{2} \varphi(1-\varphi) +   B_e  \frac{\varphi^2}{2} (1- \sqrt{\varphi})  + B_i \frac{\theta^{3/2}}{2}\biggl(1 - (1-\varphi)^{5/2} -\frac{1}{2} \varphi(5-3\varphi)\biggr)
\end{eqnarray}

where again $\varphi = \frac{\phi}{\psi}$, which coincide with (44) and (45) of \cite{S00}, respectively. They reduce in case of $B_i=0$ to our (26),(28) of Sect.IV.1. With (51), (52) we now have a situation in which all three trapping scenarios $(k_0 ^ 2, B_e, B_i )$ contribute simultaneously.\\
From (52) it follows by differentiation

\begin{eqnarray}\nonumber
-\mathcal {V}'(\phi)/ \psi^3= \frac{k_0^2}{2}(1- 2 \varphi) +   B_e \varphi (1- \frac{5}{4}\sqrt{\varphi})  + B_i \frac{5\theta^{3/2}}{4}\biggl((1-\varphi)^{3/2} -(1-\frac{6}{5}\varphi)\biggr).
\end{eqnarray}

It then follows that at $\phi=0$ it holds $-\mathcal {V}'(0) \simeq 0$ and at $\phi=\psi$: $-\mathcal {V}'(\psi) \simeq -2k_0^2- B_e + B_i \theta^{3/2}$. A positively polarized hole is then given by $k_0^2=0$ and $B_e - B_i \theta^{3/2} > 0$. This yields an extension of our previous elementary $\sech^4(x)$ solitary electron hole mode (Sect.III.2) by the $\alpha$-trapping scenario of ions.\\
If we only limit ourselves to ion trapping, a positively polarized ion hole is provided by  $k_0^2=0$, $B_e =0$ and  

\begin{eqnarray}
-\mathcal {V}(\phi)/ \psi^2= B_i \frac{\theta^{3/2}}{2}\bigg(1 - (1-\varphi)^{5/2} -\frac{ \varphi}{2} (5-3\varphi)\biggr)= B_i \frac{\theta^{3/2}}{2} (1-\varphi) \bigg[1-\frac{3\varphi}{2} - (1-\varphi)^{3/2} \bigg].
\end{eqnarray}
Together with the corresponding NDR, this forms the basis for $ \bf {positively} $ - polarized ion holes, a previously unknown and unexplored area. Some more details could be further explored, such as the corresponding Schamel-type evolution equation, but we'd like to leave that up to the reader and/or later generations.\\
Finally we just want to show that the present formalism includes the usual (negatively) polarized ion hole.\\
This is achieved by setting $k_{0-}^2:=2k_0^2 + B_e - B_i \theta^{3/2}\equiv 0$  (see (48) of \cite{S00}).
In case of $B_i=0$ we get back the negatively polarized SEH (27), and for $B_e=0$ we obtain
$-\mathcal {V}(\phi)/ \psi^2= B_i \frac{\theta^{3/2}}{2}(1-\varphi)^2 \bigg(1 - \sqrt{1-\varphi}\bigg)$.
This is obviously our familiar negatively polarized ion hole (see (7), (8) of \cite{SB80}) which holds for the dependent variable $\hat\varphi:=\varphi - 1 \le 0$.

\section{ Stability}

For understandable reasons, the last word cannot be said on the stability of these structures. The dynamics that are triggered by perturbations depend too much on what is happening in the resonance region for a general, conclusive statement to be made. This applies all the more to studies in which such an equilibrium solution was not available.
Since the mathematical endeavor turns out to be too complex, we can only outline its general properties, but solve it in the case of a single wave.\\

To make the analysis as transparent as possible, let's focus on trapping of electrons in its simplest, nontrivial version and on  $ \theta = 0 $ corresponding to immobile ions ($ n_i = 1 $).\\
By the ansatz 

\begin{eqnarray}\nonumber
f_e(x,v,t)=f_{0e}(\varepsilon) + f_1(x,v) e^{-i\omega t} + c. c.\\
\phi(x,t)= \phi_0(x) + \phi_1(x) e^{-i\omega t} + c. c.
\end{eqnarray}
where $f_{0e}(\varepsilon)$ and $\phi_0(x)$ are our equilibrium functions, we get by linearizing the VP system and by using the  integration technique along unperturbed orbits (characteristics) of \cite{LeSy79,S82a} a non-local eigenvalue problem for ($\omega,\phi_1(x)$) of the following form:

\begin{eqnarray}
\Lambda \phi_1(x):= \biggl(\partial_x ^2 + \mathcal V''(\phi_0(x)) \biggr) \phi_1(x)  =  \int dv \partial_{\varepsilon}f_{0e} \sum_{n=0}^{n=\infty}\biggl(\frac{-i}{\omega} v \partial_{x}\biggr)^n \phi_1(x),
\end{eqnarray}
which is (26) of \cite{S18}. We restrict the analysis to the cnoidal electron hole case of Sect.IV.1, in which $\mathcal V(\phi_0)$ is represented by (26), i. e. we ignore for convenience all the other electron trapping scenarios. It then holds to first order in $S=\frac{4B}{k_0^2}$\\
$ \mathcal V''(\phi_0)= k_0^2 - B \biggl(1-\frac{15}{8} \sqrt{\frac{\phi_0}{\psi}}\biggr)=k_0^2\biggl(1 -\frac{S}{4}[1-\frac{15}{8}\cos\frac{k_0 x}{2}]\biggr)$, where in the last step we used $S<<1$ .\\
The eigenvalue problem (55) then becomes 
\begin{eqnarray}
\biggl(\partial_x ^2 + k_0^2\biggr) \phi_1  - \frac{S}{4}k_0^2\biggl(1-\frac{15}{8} \cos\frac{k_{0}x}{2} \biggr) \phi_1(x)  =  \int dv \partial_{\varepsilon}f_{0e} \sum_{n=0}^{n=\infty}\biggl(\frac{-i}{\omega} v \partial_{x}\biggr)^n \phi_1(x),
\end{eqnarray}

In the harmonic (single) wave limit, when $S=0$ and $\phi_1 \sim e^{ikx} + c.c.$, it reduces to the algebraic equation

\begin{eqnarray}
(-k^2 + k_0^2)= \int dv \partial_{\varepsilon}f_{0e}\sum_{n=0}^{n=\infty}\biggl(\frac{kv}{\omega}\biggr)^n=\sum_{n=0}^{n=\infty}\biggl(\frac{k}{\omega}\biggr)^n \mathcal M_n(\phi_0),
\end{eqnarray}
which is (27) of \cite{S18}. The moment $\mathcal M_n(\phi_0)$ is defined by $\mathcal M_n(\phi_0)=\int dv v^n \partial_{\varepsilon}f_{0e}$ and it holds the recursion formula $\mathcal M_{n+2}'(\phi_0)=(n+1)\mathcal M_n(\phi_0)$, which is (29) of \cite{S18}. From $\mathcal M_2(\phi_0)=-\int dv f_{0e}= -n_{e0}(\phi_0)$ and the recursion formula we obtain $\mathcal M_0(\phi_0)= -n_{e0}'(\phi_0) =\frac{1}{2}Z_r'(\frac{\tilde v_D}{\sqrt 2}) + O(\sqrt\phi_0, S)$. Whereas the odd moments vanish, the other even moments are of $ O(\phi_0)$ or of higher order. We then get to lowest order 
$-k^2 + k_0^2 =\mathcal M_0+(\frac{k}{\omega})^2\mathcal M_2=\frac{1}{2}Z_r'(\frac{\tilde v_D}{\sqrt 2})-(\frac{k}{\omega})^2$, which is (31) of \cite{S18}. Application of the NDR (11) with $\theta=0=\Gamma$ we hence find $\omega=1$.\\
The perturbed eigenmode is an undamped, infinite wavelength  Langmuir mode (or a pure plasma oscillation, respectively, \cite{SMS17}), being independent of $\tilde v_D$.
\\
A harmonic (single-wave) EH is therefore marginally stable, no matter how strong $ \tilde v_D $ is, a result which contradicts Landau theory, where marginal stability holds at threshold  $ v_D=  v_D*$ only. \\

 Sentences like "the single-wave model ... describes the behavior near the threshold and subsequent nonlinear evolution of unstable plasma waves"  \cite {BMT13} rest on the unproven ad hoc assumption of the validity of the linear Vlasov concept which can, however, not be retained. The underlying single-wave model therein is too simple and simply not applicable. These authors underestimate the effectiveness and need of particle trapping in the real world of coherent structures, an unmistakably nonlinear effect. A similar misunderstanding of the importance of trapped particles is encountered in \cite{VAL12} (see also \cite{S13,Dec21}). Here the authors did not realize that their analysis is based on nonlinearly fake modes. As in the case of the "Thumb-Teardrop DR", see Section III.1, the on- and off-dispersion modes only make sense and become real modes when trapping is built in. The fact that they are also encountered as linear Vlasov modes is correct, but overlooks the fact that the associated distribution functions are no longer valid nonlinearly, as they should.\\

In contrast to the currently favored wave theory, which is based on Landau's analysis and is vehemently defended by its protagonists, a nonlinearly permitted single wave  is $\bf{unconditionally}$ $\bf{ marginally}$ $\bf{ stable}$. \\

For mobile ions we get as an extension of (57):
\begin{eqnarray}
(-k^2 + k_0^2)= \sum_{n=0}^{n=\infty}\biggl(\frac{k}{\omega}\biggr)^n \biggl[\mathcal M_n(\phi_0) +\frac{\theta}{\mu^n}\mathcal M_n^{i}(\phi_0)\biggr],
\end{eqnarray}
with $\mu=\sqrt \frac{\theta}{\delta}$ and a corresponding recursion formula for $\mathcal M_n^{i}(\phi_0)$. It is found that  $\mathcal M_0^{i}(\phi_0)= \frac{1}{\theta}n_{i0}'(\phi_0) =\frac{1}{2}Z_r'(\frac{u_0}{\sqrt 2}) + O(\phi_0)$ and  $\mathcal M_2^{i}(\phi_0)= -n_{i0}(\phi_0)$ such that to lowest order we have: 
$-k^2 + k_0^2 =\frac{1}{2}Z_r'(\frac{\tilde v_D}{\sqrt 2})-(\frac{k}{\omega})^2 + \theta\biggl[\frac{1}{2}Z_r'(\frac{u_0}{\sqrt 2})-(\frac{k}{\omega \mu})^2\biggr]$, from which follows $\omega=\sqrt{1 + \delta}$.
This is the undamped Langmuir mode with infinite wavelength corrected for the mass ratio as it should be.\\

 In the opposite limit of a maximum distortion, $k_0 \rightarrow 0$ and $S \rightarrow \infty$, which is the solitary wave limit, this linear stability problem was attacked by \cite{S82, Hu18}. A solution was obtained by an artificial truncation of the series in (55) at $n=2$, the so-called fluid limit \cite{LeSy79}, and by a subsequent representation of $\phi_1(x)$ in terms of the eigenstates of the $\Lambda$ operator, which is the operator on the left hand side of (55). The result of a longitudinal stability and a transversal instability has however to be questioned because there are hints \cite{S86} that this artificial truncation of the series cannot be justified.\\
This, as well as the stability problem for any $ S $ and for all the other trapping scenarios, is hence a great challenge and can occupy many generations.\\

\section{ Negative energy states and spontaneous hole acceleration}

Another important, if not the most important, aspect is the fact that the total energy of a plasma can be less than that of the undisturbed plasma due to the presence of a hole.This implies that when this state is approached, for example by achieving a higher phase velocity through acceleration, free energy is available which can be the source for the excitation of other modes and thus for a higher degree of intermittent plasma turbulence. As an example I refer to the simulations of 
\cite{SMS17,MSS18,MSS20} and the corresponding video [https://youtu.be/-nxIokKORwU] (with gratitude to my coauthors Mandal and Sharma \cite{SMS17,MSS18,MSS20,SMS20a,SMS20b}). 
We see in this video an acceleration that is particularly efficient when $ \theta> 1 $, the existence condition for ion acoustic waves. This transition to a higher phase velocity must be a transient, unsteady process, because the NDR (5)  written as $ -\frac {1}{2} Z_r '(\frac {u_0} {\sqrt 2})=c $ exhibits  as a stationary NDR a forbidden area or a gap between the slow and the fast branch when $-0.285 < c< 0$  (see section III.B of \cite {MSS20} or section 4.1 of \cite {MSS18}). For energetic reasons the continuous process of ion sound wave emission during acceleration must hence be associated with a reduction in the hole energy. This process is somewhat similar to the radiation from KdV solitons and Langmuir solitons investigated  by Karpman and coworkers (\cite{Karp95} and references therein).\\ 

 Experimentally a spontaneous acceleration of periodic ion holes was detected by \cite{FKPS01} in a double plasma device.\\

Another process is worth mentioning, the excitation of an energetic plasma oscillation, as can be seen in the behavior of $ \phi (x, t) $  of the simulation. As explained in \cite {SMS17}, this is a relic of two counter-propagating Langmuir waves of the same intensity which can already be understood by a linear fluid approach of the electrons. This plasma oscillation carries most of the excess energy added to the plasma by the initial disturbance.\\

There is therefore great interest in learning more about the energy associated with a hole which is our last topic.
It was developed in a series of papers \cite{S00, GS02, GLS02, LS05, DBS18}, to which we refer for a more intensive evaluation. The total energy density $w$ of a plasma  that is structurally excited by an equilibrium hole is given in the laboratory system by
\begin{eqnarray}
w =\frac{1}{4L}\int_{-L}^{+L}dx'\biggl[\int_{-\infty}^{+\infty}dv v^2 f_e(x',v) +\frac{1}{\theta}\int_{-\infty}^{+\infty}du u^2 f_i(x',u) + \phi'(x')^2 \biggr]
\end{eqnarray}

where $x'=x-v_0t$ and a stationary structure of periodicity $2L$ ($L\rightarrow \infty$ for solitary holes) is assumed.
The distributions are given by (1) for electrons and by (48) for ions, respectively.\\
Using a straightforward calculation (\cite{S00,GS02,GLS02}) $w$ is found to be
\begin{eqnarray}\nonumber
w =\frac{1}{2}\biggl[(1+\frac{k_0^2\psi}{2})(1+v_D^2) + \int_{0}^{\psi} n_i(\phi)d\phi +\frac{1+K_i}{\theta}\biggr] 
 + \frac{1}{4L}\int_{-L}^{+L}\biggl(v_0^2[n_e(\phi)- n_e(0)] + \frac{u_0^2}{\theta}[n_i(\phi)- n_i(\psi)]\biggr)dx\\
 -\frac{3}{4L} \int_{-L}^{+L}\mathcal V(\phi)dx,
\end{eqnarray}
which is (67) of \cite{LS05} (in which $K$ stands for $k_0^2\psi/2$ and $A$ for $K_i$). This expression reduces to $w_H:=\frac{1}{2}(1 + v_D^2 +\frac{1}{\theta})$ in the structureless, homogeneous plasma limit $\psi \rightarrow 0$. An appropriate renormalization of the electron quantities, which takes into account that $\frac{1}{2L}\int_{-L}^{+L}n_e dx = 1 + \tilde \sigma$ can deviate from unity ($|\tilde \sigma|<< 1$) yields $w_S:=(1-\tilde \sigma)w$,
which is (69) of \cite{LS05}.\\
Defining finally $\Delta w$ by $\Delta w:=w_S - w_H$ we arrive at the energy (density) difference provided by the structure.\\
To simplify the further discussion we only consider the $B_e$ ($\equiv B$) and $B_i$ trapping scenarios (corresponding to (20a,b) of \cite{S00} and get (see (73) of \cite{LS05}

\begin{eqnarray}\nonumber
\Delta w = \psi \biggl[\frac{1}{2}\biggl(1+(\frac{k_0^2}{2} - \hat\sigma)(1+v_D^2) + \frac{\hat K_i - \hat \sigma}{\theta}\biggr) + 
  \frac{v_0^2}{4L}\int_{-L}^{+L}[-\frac{1}{2}Z_r'(\frac{\tilde v_D}{\sqrt 2}) \varphi - \frac{5}{4}B_e \varphi^{3/2}]dx \\
 +\frac{u_0^2}{4L\theta} \int_{-L}^{+L}[-\frac{\theta}{2} Z_r'(\frac{u_0}{\sqrt 2}) (1-\varphi) - \frac{5}{4} B_i(1-\varphi)^{3/2}]dx\biggr]                 
\end{eqnarray}
with $\hat\sigma:=\frac{\tilde\sigma}{\psi}= \frac{k_0^2}{2}-\frac{1}{2}Z_r'(\frac{\tilde v_D}{\sqrt 2}) \frac{1}{2L}\int_{-L}^{+L}\varphi(x) dx - \frac{5}{4} B_e \frac{1}{2L}\int_{-L}^{+L}\varphi(x)^{3/2} dx$      and           
$\hat K_i:=\frac{K_i}{\psi}= \frac{\theta}{2} Z_r'(\frac{u_0}{\sqrt 2}) +\frac{5}{4}B_i$.\\
We note that in (61) terms of $O(\psi^2)$ have already been neglected which means that the field energy term, the last term in (60), does no longer contribute since it is $O(\psi^2)$.\\
In the solitary, positively polarized electron hole limit ($k_0^2 \rightarrow 0$) (61) becomes (see also (77) of \cite{LS05})
\begin{eqnarray}
\Delta w =\frac{\psi}{2}\biggl(1+[\frac{1}{2} Z_r'(\frac{u_0}{\sqrt 2}) + \frac{5}{4\theta} B_i](1-u_0^2) \biggr)               
\end{eqnarray}
which  extends (8) of \cite{GS02}.

We learn that $ \Delta w $ is only influenced by the ion response and by the ion trapping scenario $ B_i $ whereas $ B_e $ only participates implicitly through the back door via the NDR. \\
An inspection of (62) shows that in order to change the sign of $ \Delta w $, $u_0$ has to be larger than some $u_0^*$ which is defined by $ \Delta w(u_0^*)=0 $ and which only depends on  $B_i/\theta$. In case of $B_i=0$ it is given by
 $u_0^*=2.124$. If $B_i>0$  $u_0^*$ will grow monotonically from 1 ($\theta=0$) to 2.124 ($\theta =\infty$).
The dependenc of $u_0^*$ is plotted in Fig.8 of \cite{LS05} whereas the region of $\Delta w <0$ in the ($v_D,\theta$) plane is exposed in Fig.10 (for $B_i=0$).\\
This brings us to three basic properties of a structurally excited plasma:\\

1] $w_S - w_H=\Delta w= O(\psi)$, the difference in the total energy density is $ O(\psi)$ rather than  $ O(\psi^2)$ as found e.g. by standard linear wave analyses (see Appendix A5, \cite{KO58, Gardner63,MP94}). The influence of a coherent structure on the energy budget is hence much stronger than predicted by linearly based concepts.\\

2] By acceleration, a hole can penetrate into areas with negative energy and thus release energy that the plasma can use to generate further waves and hence to increase  the level of intermittent turbulence.\\

3] Due to the multitude of different trapping scenarios, there is a vast, untapped field that many generations of plasma theorists can still benefit from.

\section{ Two related topics: anomalous transport and holes in synchrotrons}

X.1 Coarse grained distributions and anomalous resistivity\\

The point to be addressed is that the current equilibria still have a weak singularity of cusp type in phase space. At least the free part of the distributions $f_e$ (and $f_i$) exhibits an infinite slope at the separatrix, $|\partial_v f_e(\varepsilon)| \sim \frac{1}{|\varepsilon|}$ as $|\varepsilon|\rightarrow 0$ (and similarly for the ions), which is an integrable singularity.
In other words collisional aspects enter near the separatrix and higher moments of the BBGKY hierarchy have to be considered in this region \cite{KrallTr73}.\\
Numerically this problem was attacked by the authors \cite{KS96a,KS96b, SK96, LS05} who added a Lorentz-collision operator, $\nu_e\partial_v[\partial_v + ( v- \tilde v_D)]f_e$, on the right hand side of the kinetic equation where $\nu_e$ is the electron collision frequency. By the inclusion of a homogeneous electric field on the left hand side, $E_0=-\nu_e v_D$, they got dissipative structural equilibria of the so-extended Vlasov-Fokker-Planck-Poisson system. We refer to Fig.6 of \cite{KS96b} for fluid ions and to Sect.2.5 of \cite{LS05} for kinetic ions.\\
The distribution on the separatrix was smooth in all cases, which corresponded to a coarse-grained distribution, and a single-humped curve could be obtained in the $ (\nu_e, v_D) $ space, below which dissipative hole equilibria were time-asymptotically established (see Fig.9 of \cite{KS96b} and Fig.23 of \cite{LS05}). In turn, ion mobility played a decisive role for the existence of these dissipative structural equilibria. Accordingly, an anomalous resistivity (or conductivity, respectively) could be presented, Fig.10  and equation (5.13) of \cite{KS96b}, which was determined and controlled by the surviving hole structure.\\
Holes therefore play a crucial role in anomalous transport and it is expected that this area will receive more attention in the future.\\
This approach to intermittent plasma turbulence with surviving structures and coarse grained distributions is supported by another investigation \cite{SL05}. In this numerical study of a current-carrying, subcritical pair plasma ($ v_D = 2.0 < v_D * = 2.6$), the nonlinear growth (rather than damping!) of holes in the positive species, which were initially triggered by tiny seeds, could be demonstrated. After saturation a new, steady-state, collisionless, intermittent plasma turbulence state is approached with persistent, albeit somewhat less energetic, holes in it.\\
We note that in VP simulations and/or PIC simulations such a coarse graining is automatically involved by the artificial phase space diffusion, especially at the separatrix, that is triggered by the numerically necessary discretization of space.
We may therefore speculate that the current collisionfree equilibria could be of fundamental importance for the derivation of improved and specially adapted collision operators in weakly collisionfree plasmas such as fusion or space plasmas.\\

X.2 Solitary structures on hadron beams in synchrotrons \\

Another field of application of our theory of VP structures are the structures that were measured on continuous (coasting)  as well as bunched  particle beams in circular accelerators e.g. at RHIC \cite{Blas03}. Our theory, which has been developed over almost a decade  \cite{LS05, S97, S98,  SF00, GSF00, Blas04, SL04} combined with the present experience, does predict\\
 a)  no threshold for nonlinear structure formation in case of a coherent initial fluctuation spectrum provided that the system is above the transition energy and\\
 b) the existence of long-living stabilized structures that belong to a continuous rather than a discrete spectrum. 

The loss of Landau damping, which is seen within these accelerators when at higher intensities the collective frequencies lie outside the incoherent spectrum, is, however, wrongly explained by the beam community (see e.g. \cite{Burov21}) through a tune shift in the sense  that there are no longer any particles that can interact resonantly with the wave structure, instead of correctly interpreting this phenomenon through a loss of Vlasov linearity.\\
 In addition, another incorrect interpretation is given (see e.g. \cite{Karpov21}) that the observed continuous spectrum after saturation is of the discrete van Kampen type, instead of recognizing that the linearity no longer applies and continuous spectra of the type presented in this work should instead be implied.\\ 
We conclude with the expectation that particle trapping, the associated nonlinearities and the various trapping scenarios will certainly find their way into beam physics in the future as well. \\

\section{Summary and Conclusions}
The aim of this review was to prove  that $\bf{coherent}$ $ \bf{ electrostatic}$ $\bf{ structures}$ are due to the (trapping) nonlinearity of the Vlasov-Poisson system. By comparison of exact nonlinear with approximative linear wave
solutions of current-carrying plasmas with drift velocity $v_D$ two fundamentally new results could be obtained:\\
(i) nonlinearly proper single (harmonic) waves are linearly marginally stable independent of $v_D$ (see harmonic mode, section III.1 and stability, section VIII) and\\
(ii) a  spontaneous acceleration of a tiny hole is observed, triggered by a gap in the velocity caused by the nonlinear dispersion relation and accompanied by its subsequent approach to a more negative energy state (see section IX).\\
This necessarily  implies that any linear treatment of this system, such as that of Landau or van Kampen (Case), turns out unsuitable for approximating them. Their diversity and mathematical subtleties are a direct result of the various trapping scenarios that are caused by the resonant wave-particle interaction, a problem which is mathematically known to be nonlinear and nonintegrable. \\
With this work we have thereby laid the basis for a comprehensive description of coherent patterns in current-driven, noisy plasmas, which brings structure and order into the manifold of hole equilibria. They originate from six different electron trapping scenarios that reflect the chaotic single-particle trajectories in the vicinity of the resonance. Thirty-one qualitatively different solutions for  the electrostatic wave potential $\phi(x)$ can be composed which are achieved by all possible combinations of five elementary modes, a manifold that includes thirty different modes of solitary wave character, namely fifteen  of positve and fifteen of negative polarity. These elementary modes are: $\sin (x), \sech^4(x), e^{-x^2}, e^{-sinh^2(x)}, \sech^2(x)$. When, in addition,  ion trapping effects are taken into account, the result is a three-digit number of modes around five hundred.

 In general, however, two combinations of them are already sufficient to prevent the electric wave potential $ \phi (x) $ from being disclosed, i.e. it can no longer be described mathematically by known functions.  A distinction between these structures is still possible, however, on the level of the pseudo-potential $ \mathcal V (\phi) $, which is the central variable in the present  theory, leading to a $\bf{nonlinear}$ version of the linear $\bf{superposition}$ $\bf{ principle}$.

Its utmost generality has the consequence that earlier investigations presented so far are either included as special cases or appear in a  corrected, updated form. It offers a profound foundation since it provides a  mathematically precise in-depth microscopic derivation. This is in contrast to many studies presented so far, in which such a deepened phase space study is missing. A well-founded study is absent either because for example, the densities are simply given as functions of $ \phi $ rather than being derived, or the studies are wrongly performed without realization that the phase velocity is a necessary part of a complete wave theory. In addition, an intrinsically microscopic process can in generality not be adequately addressed by prescribing a macroscopic potential $ \phi (x) $ as done by the BGK method. 

The shortcomings of current methods such as the BGK method \cite{BGK} and especially those based on the linear Vlasov theory (van Kampen, Landau) have hence been identified  and  corrected in favor of the present theory. In noisy plasmas, which are characterized by localized fluctuation nuclei or eddy-like seeds of non-topological character, the linear Landau theory fails, contrary to the popular opinion. The linear Landau theory is undermined by particle trapping, the sibling of coherence, and fails to correctly contribute to the formation of coherent patterns in realistic plasmas because of its resonantly  inconsistent distributions. 

The present status report was limited to continuous distributions, to a Maxwellian background plasma, to non-relativistic electrons and to non-magnetized plasmas. Various extensions were carried out in the literature with regard to discontinuous distributions by \cite {S15,SDB18,DBS18}, to $ \kappa $ or Fermi-Dirac distributions by \cite {Tribeche12, Haas21} and \cite{SE16}, respectively, to relativistic electrons by \cite {ES05,ES06} and to magnetized plasmas by \cite {Jov02}.

As pointed out, however, a continuation is still called for because of the cusp singularity at the separatrix, which is  inherent in all solutions and which requires an extension of the kinetic Vlasov-Poisson description by incorporation of the higher moments of the BBGKY-hierarchy. This is definitely a new adventure awaiting us in which the mutually dependent chaotic particle behavior and the pair correlations at resonance enter into a liaison with an open outcome, especially for  intermittent plasma turbulence and anomalous transport. Such an implementation is still pending and will certainly keep many generations busy.\\
Other non-trivial challenges are 3D generalization of holes \cite {Jov02, Chen02}
or  the incorporation of a second potential, the vector potential, in order to take into account the additional magnetic island formation, the latter being required, for example, for the propagation of coherent kinetic Alfv$\acute e$n waves \cite {Karim13}. \\
The proximity to the incompressible, non-viscous shear flow in 2D (Rayleigh problem) and more generally to the general fluid theory \cite {S12} yields a further application. The "puffs" observed in long pipe flows at high Reynolds numbers \cite {Hof08} seem to be closely related to the mentioned phase of hole acceleration and sound emission. The latter is caused by the approach of the system to an energetically lower hole state. This similarity of the coherent vortex dynamics in phase and real space physics, respectively, should definitely deserve further attention in future investigations. \\
In addition, this analysis can be extended point by point to include collisionfree shocks or double layers. This is achieved in the solitary wave case of $k_0=0$ by the further constraint:  $ \mathcal {V}'(\psi)=0$ or $n_e(\psi)=n_i(\psi)$ with $n_e(\phi)$ from (2) and $n_i(\phi)$ from (49). Accordingly, the variety of shock solutions is unlimited, as in the solitary wave case, being determined by particle trapping.\\
And of course finite amplitude extensions like that  in \cite {S72, SB80, BS81, SB83, S82, S86} remain a gigantic challenge for future generations.\\
 In the case of a  bump-in-tail driven plasma such a comparison is still missing because, to my knowledge, no suitable nonlinear solution for comparison is currently available. However, there is no obvious reason why linear wave theory should surprisingly be applicable to coherent structures in this second case, as is practically assumed in the (especially experimental) literature.\\

Given the relatively limited progress that the plasma community has made in describing coherent phase-space structures over the past half-century, it would be desirable if the above article could help point the way in a new direction. Not only the insistence on the linear Vlasov description (Landau, van Kampen) set the wrong course, also the uncritical application of the nonlinear BGK method got stuck in a preliminary incomplete phase. As explained in the present article both positions are outdated or, at least, need to be improved. With the complexity of both $\phi$ and $v_0$, paired with the chaotic  single particle trajectories and associated trapping scenarios as well as the necessary extension through pair and higher order correlations, the theory of phase-space structures in noisy, collision-free driven plasmas is therefore more in its infancy than a closed topic.\\

We finish by concluding that Landau's theory is inapplicable to  driven plasmas with a non-negligible background of fluctuations. To describe the 1D instability of those plasmas that violate the topological constraint of Landau's theory, one has to solve the full unabridged Vlasov equation, i.e. without linearization.
The pattern formation induced by minute seed fluctuations in collision-free plasmas is thus intrinsically nonlinear since it is governed by particle trapping.
\\

ACKNOWLEDGEMENTS\\

I would like to thank Sarbeswar Bujarbarua for his contributions on holes with finite amplitude and double layers, J$\ddot u $rgen Korn for his contributions and deep insights into weakly collisional and ionic effects on electron holes and anomalous transport (his spectral Fourier-Hermite Vlasov-Fokker-Planck-Poisson Code being still the measure of all things), Alejandro Luque for his contributions on weakly collisional effects on holes and anomalous transport as well as on quantum effects and structure formation in bunched and coasting beams in accelerators, Florian Bauer, Prathana Borah, Nilakshi Das, Bengt Eliasson, Renato Fedele, Kalyan Goswami, Jean-Matthias Grie$\ss $meier, Vladimir Karpman and Dusan Jovanovich for their valuable contributions to various hole-related issues, and Debraj Mandal and Devendra Sharma for their series of insightful electron-hole simulations. My thanks also go to Prof. Friedrich Busse for valuable discussions and diverse support. And last but not least, I am very grateful to my wife Helga for keeping my back free over the many decades of our life together. \\

APPENDIX A  Annotated list of false statements\\

We present below an annotated list of common mistakes and pitfalls that occur frequently in the literature regarding phase correlated structures as they are based either on linear wave theory or on historical but outdated expressions.\\

A1 The threshold or onset of destabilization of a current-driven plasma is given by Landau's theory\\

The untenability of this headline becomes clear after reading and understanding this article. With a harmonic  linear Vlasov solution one is far away from a proper single wave  Vlasov solution, since trapped particles are a necessary component of  undamped coherent plasma waves. Landau's theory at the threshold satisfies the linear but not the nonlinear Vlasov equation as it should be. In addition, a correct single wave is marginally stable for all drift speeds in a current-carrying plasma and cannot be classified as damped and growing. \\
The correct treatment of coherent, low amplitude equilibrium structures will hence remedy this false headline. Its failure arises from the fact that in real, i.e. noisy plasmas with a given level of coherent initial  fluctuations  $ f_1$ the criterion for linearity: $|\partial_v f_1|<<|\partial _v f_0|$ is locally violated almost everywhere in phase space. There is therefore no need to search for a linear instability mechanism, such as a two-stream instability, to let  such structures being existent, although of course stronger drifts naturally facilitate their excitation and growth. A new destabilization mechanism within the full Vlasov approach needs to be invoked to cope with the changed start of the evolution but that is not yet in sight. For more details, see Sect. VIII.
\newline
Conceptually analogous statements are the following three:\\

A2 A drift velocity $v_D$ is subcritical, critical  or supercritical\\

This expression is again meaningless in a noisy plasma situation since for initial, seed-like fluctuations Landau's theory does not hold and a critical $v_D*$ does not exist. Hole structures don't care about $v_D*$. If there is an influence, then it is nonlinear and complements that of the other system parameters. The use of these adjectives should therefore be treated with caution and better avoided or at least mentioned in quotation marks such as "subcritical". In a plasma with a fluctuating background there is no longer a discrete cut for $ v _D $ in the form of $ v _D * $, which indicates the beginning of the instability \cite {MSS22}. \\

A3  The thumb-tear drop relation is a linear dispersion relation and is solved by the electron acoustic wave\\

Stationary monochromatic (single) waves with small amplitudes are determined by the so-called thumb-tear drop dispersion relation, which can be derived by both, the linear and the non-linear Vlasov approach. However, this macroscopic relation misses validity linearly, since in this approach the resonance particle region is incorrectly treated  at least for moderate velocities. Only in the fluid limit, i.e. for high phase velocities in each species (Langmuir for $ v_0 >> 1 $ and ion acoustics for $ u_0 >> 1 $), it remains linearly acceptable, but in this case no resonance effects need to be considered. Microscopically, for waves with velocities  in the thermal range when resonance effects play a major role, only the non-linear Vlasov approach is suitable for justifying such coherent wave equilibria.\\
One solution for small $k<<1$ is the slow electron acoustic wave (SEAW), $\omega/k =1.307$, which is called slow to distinguish it from the Gould-Trivelpiece acoustic mode (in case of a transverse boundary) \cite{S79}. This is analogous to the ionic case, where the former mode is called slow  ion acoustic wave to distinguish it from the ordinary ion acoustic wave.\\

A4 Expressions such as Landau resonance, Landau contour, nonlinear frequency shift or group velocity are terms transferable to coherent hole structures\\

The Landau contour describes the route in the complex velocity integration in a Fourier-Laplace treatment of the linear Vlasov equation, how the pole must be bypassed in order to obtain  time-asymptotically a dispersion relation with a discrete solution $\omega(k)$ \cite{Landau46, KrallTr73}. Since this theory does not hold in the stationary (or marginally stable) case\\ Im $\omega (k) = 0 $, there is due to continuity no reason why it should work for Im $ \omega (k) \ne 0 $.

It is therefore by no means surprising when authors in \cite{Be09,Be10}  find strong deviations of the $ group$ $velocity $ as a result of particle trapping and conclude that the group velocity $v_g$  of an essentially undamped wave, calculated by using the very definition of Rayleigh   \cite{Ham39,Brill60,Li65,Whit74,DeMo95}       is found to significantly differ from $\partial \omega/ \partial k$  or that surprisingly enough the main nonlinear change in $v_g$ occurs once the wave is effectively undamped."

These terms hence come from a linear body of thought and are of little application potential. It is correct, however, that coherent  hole modes belong to a continuum rather than to  a discrete class of solutions and that their time dependency is no longer determined by Landau resonances. The term nonlinear frequency shift is also less useful as it suggests that the small amplitude modes are linear modes.  The correct speed of a modulated coherent periodic hole wavelet is given by the phase speed $v_0$ since in each individual hump the trapping is adjusted such as to make the speed of the envelope valid for each hump (see \cite{Be09, Be10, LeMi13, S12}).\\

A5 The energy density of a hole structure of strength $\psi$ is $O(\psi^2)$\\

With a linear structure of the amplitude $ \psi $, the energy density of the plasma changes by a quantity of the order $ O (\psi ^ 2) $. No matter whether the plasma is treated as a dielectricum \cite{KO58, Gardner63} or refined by treating resonant particles more carefully linearly \cite{MP94} the result is $O(\psi^2)$. This is in strong contrast to a proper nonlinear treatment \cite{S00,GS02,GLS02,LS05, S12, DBS18} since the result is  $O(\psi)$ (for more details see Sect.IX). This not only implies that the effect appears one  order earlier in an expansion procedure and is hence stronger but also that negative energy states occur more frequently. This is of importance since in a  Cauchy initial value problem triggered by seeds or coherent initial fluctuations these states are time-asymptotic attractors \cite{MSS18} that contribute effectively to the intermittent plasma transport. For more details we refer to Sect. IV.\\

A6  Stationary electrostatic wave structures are Bernstein, Greene, and Kruskal (BGK) modes\\

This is an at least inaccurate statement for three reasons. First, a BGK mode, i.e.  a structure obtained by the BGK method, only accounts for the shape $ \phi (x) $, but leaves $ v_0 $, the phase velocity, indefinite. It is therefore incomplete.
Second, in a typical case $\phi(x)$ can not be described  mathematically by known functions. It is undisclosed due to a manifold of potential  trapping scenarios, as the present paper shows. To get analytically the searched distribution of trapped particles, $f_{et}$,  the concrete form of $\phi(x)$ is however needed. Third, $ f_ {et} $ can be non-physical, for example negative in parts of the phase space, which is physically unacceptable because there are no negative probabilities.\\
Nevertheless, the BGK method can be valuable and symbiotic with the Schamel method, as it may suggest  new particle trapping processes. An example is the Gaussian profile which resulted in a logarithmic trapping scenario \cite{S71, Kras97, Musch99, Chen02}.
Valid therefore is that only in combination  with the Schamel method a complete access to coherent equilibrium structures can be achieved. \\

A7  BGK modes become in the small amplitude limit van Kampen modes and are the most general\\

It is generally claimed that BGK modes reduce in the small amplitude approximation to what is known as van Kampen modes. This is false, however, since such a transition from a nonlinear to a linear mode does not take place, even in the infinitesimal amplitude limit. A harmonic hole equilibrium of the Vlasvov-Poisson system that is correctly described by the Schamel method, inclusiveley its phase velocity, shows that nonlinearity persists in all stages of this limiting process. The region of trapped particles never disappears and there is no critical point at or below which the trapped distribution collapses into a $\delta$-function or other linear functions. Another indication that this claim is incorrect is that such nonlinear modes are unconditionally marginal stable in current-carrying plasmas independent of the drift velocity between electrons and ions (see Sect.VIII). Landau's theory of damping and growth, respectively, as a linear wave theory is obviously inapplicable in our case of coherency. \\
The privilege of being the most general method is therefore reserved for the Schamel method, since it is as general as the BGK method, it is in addition complete and can also deal with undisclosed solutions. \\

A8 Holes in synchrotrons and storage rings exist above a threshold only and are van Kampen modes\\

This threshold statement is only valid if there is a certain band of incoherent sychrotron frequencies, as for the applicability of the Landau damping \cite {Burov21, Karpov21, Chin83}. For structures that arise from coherent seeds, however, there is a loss of linear Vlasov dynamics at all, which not only implies the lack of a threshold for the invalidity of Landau damping but the overall existence of hole equilibria and the exclusion of  van Kampen modes  to describe these structures  in favor of the current theory (see also Sect.X2).\\

APPENDIX B  Derivation of  (27)\\

We start with $-\mathcal {V}(\phi)= \frac{k_0^2}{2\sqrt \psi}( \phi \psi^{3/2} -3\phi^2  \psi ^{1/2}+ 2 \phi^{5/2})$ but use instead of (8) the equivalent expression \\
 $x(\phi)=\pm \int_{0} ^{\phi}\frac{dt}{\sqrt{{-2 \mathcal V(t)}}}$  which is better suited because $\phi=\psi$ occurs at infinity rather than at zero. The +(-) sign holds for $x\ge 0$ ($x<0$).  We then have with $\varphi = \phi/\psi$ :\\

$k_0x=\pm \int_{0} ^{\phi}\frac{dt}{\sqrt{{-2 \mathcal V(t)}}} = \pm  \int_{0} ^{\varphi}\frac{dt}{\sqrt{ t - 3t^2 + 2 t^{5/2}}} =  \pm 2 \int_{0} ^{\sqrt \varphi}\frac{dy}{\sqrt{ 1 - 3y^2 + 2 y^{3}}} =   \pm 2 \int_{-1} ^{\sqrt \varphi -1}\frac{ds}{s\sqrt{3+ 2s}} =\pm\frac{4}{\sqrt 3} [\tanh^{-1} (\sqrt{\frac{ 2\sqrt \varphi +1}{3}})-                                      \tanh^{-1} (\frac{1}{\sqrt{3}})] $.\\
With the abbreviations $\zeta:=\frac{\sqrt3 k_0 x}{4}$ and  $\zeta_0:= \tanh^{-1} (\frac{1}{\sqrt{3}})=0.65848     $   
we then have \\
$\varphi(\zeta)= \frac{1}{4}\bigg[ 3 \tanh^2(\zeta \pm \zeta_0) -1 \bigg]^2$\\
where again the +(-) sign holds for $\zeta \ge 0$ $ (\zeta < 0)$. This corresponds to (27) and  is our main result in this Appendix B. It is easily seen by making use of $\tanh(x+y)=\frac{\tanh(x) + \tanh(y)}{1 + \tanh(x) \tanh(y)}$ that this main formula agrees with (3.33) of \cite{KS96a} if $\zeta \ge 0$. (We mention in parenthesis that the application of (3.33) also for negative $\zeta$ would result in a small additional hump at $\zeta =-\zeta_0$ which is unphysical because it comes from a wrong handling of the equations.)\\

REFERENCES


\begin{thebibliography}{88}
\bibitem{Cit7}{Its small amplitude existence can be seen as a further hint for the \\need of analytical studies like this, especially in the perturbative trapping regime ($\beta,\gamma$)}.
\bibitem{Dec21}It is a misunderstanding of the effectiveness of particle trapping to believe Landau's onset point of instability can be justified by assuming  $f_0'(v)=0$ for $v_0-\Delta v \le v \le v_0+\Delta v$, $\Delta v \ge 0$. Even if this topological constraint applies to a given $ f_0 (v) $, which would correspond to a severe restriction of the permitted background distributions, a $ f (x, v) $ of type (1) is nevertheless essential to model the resonant region correctly in a nonlinear manner.
\end{thebibliography}
\end{document}